\documentclass[onecolumn,numbers]{els-mrw-mod} 

\usepackage{amsmath,amssymb,amsfonts,amsthm,graphicx}
\usepackage{txfonts,mathrsfs}
\usepackage{helvet}
\usepackage{tikz}
\usetikzlibrary{matrix,shapes}
\usetikzlibrary{calc}
\usetikzlibrary{arrows.meta}
\usepackage{hyperref}
\usepackage{cleveref}
\usepackage{tcolorbox}

\newtheorem{thm}{Theorem}[section]	
\newtheorem{conjecture}[theorem]{Conjecture}
\newcommand{\Graph}{\Gamma}
\newcommand{\me}{\mathsf{e}}
\newcommand{\mv}{\mathsf{v}}
\newcommand{\Eset}{\mathsf{E}}
\newcommand{\Vset}{\mathsf{V}}
\newcommand{\Bset}{\mathsf{B}}
\newcommand{\Iset}{\mathsf{I}}
\newcommand{\Lset}{\mathsf{L}}

\newcommand{\R}{\mathbb{R}}

\newcommand{\hK}{\widehat{K}}

\DeclareMathOperator\diag{diag}
\DeclareMathOperator\var{var}
\DeclareMathOperator\Hess{Hess}

\usepackage[dvipsnames]{xcolor}

\renewcommand{\term}[1]{\emph{#1}}

\begin{document}

\chapter[Quantum graph models of quantum chaos]{Quantum graph models of quantum chaos: an introduction and some recent applications}\label{chap1}

\author[1]{Gregory Berkolaiko}%
\author[2]{Sven Gnutzmann}%

%\author[1,2]{Third Author}%

\address[1]{\orgname{Texas A\& M University}, \orgdiv{Mathematics}, \orgaddress{Texas, USA}}
\address[2]{\orgname{University of Nottingham}, \orgdiv{School of Mathematical Sciences}, \orgaddress{University Park, Nottingham NG7 2RD, United Kingdom}}

\articletag{Chapter Article tagline: update of previous edition,, reprint..}

\maketitle

\begin{glossary}[Keywords]
Quantum Graphs, Quantum Chaos
\end{glossary}

%\begin{glossary}[Nomenclature]
%\begin{tabular}{@{}lp{34pc}@{}}
%AF &Assessment Factor\\
%ECHA &European Chemical Agency\\
%EPM &Equilibrium Partitioning Method Equilibrium Partitioning Method Equilibrium Partitioning Method Equilibrium\hfill\break Partitioning Method\\
%ERA &Ecological Risk Assessment\\
%HC &Hazardous Concentration\\
%\end{tabular}
%\end{glossary}

\begin{abstract}[Abstract]
  Quantum graphs are a paradigmatic model for quantum chaos as well as
  for spectral theory.  We give a concise didactical introduction to
  quantum graphs, or Schr\"odinger Hamiltonians on metric graphs, with
  a focus on results related to quantum chaos, periodic orbit theory
  and spectral theory. We summarise related seminal results, and give
  an overview of a few more recent developments.
\end{abstract}

\section{Introduction}
\label{introduction}

Quantum graphs have been a paradigm model for quantum chaos since
Kottos and Smilansky introduced the model to the field almost 30 years
ago (and coined the term `quantum graph')
\cite{KotSmi_prl97,KotSmi_ap99,KotSmi_prl00,KotSmi_jpa03}.  Variations
of the model had been used in physics before (see
\cite{Pau_jcp36, RueSch_jcp53,Ale_prb85,ExnSeb_rmp89,BulTre_jmp90}
for a selection) and it already had a history in Mathematics (see
\cite{Rot_crasp83,Rot_incol84,Bel_mmas88} for some work that are
relevant in this context).  While the term `quantum graph' is often
used by mathematicians as well the model is often referred to as
`Schr\"odinger operators' or `Laplacians' on `metric graphs'.  The
term `quantum graph' is also used in quantum information theory where
it refers to operator algebras that are based on an underlying graph
structure -- which is a different concept that will not be covered
here.

From the turn of the millennium 
quantum graphs had a surge of applications
in both mathematical spectral theory and quantum chaos, with many collaborations and
mutual inspirations which continue to this day.

The aim of the present manuscript is two-fold:

\vspace{2\lineskip}
\begin{enumerate}
\item We want to give an introduction to quantum
graphs with a perspective on applications in quantum chaos.
\item 
Give an overview of the literature for further reading and present a few recent developments
that we find interesting.
\end{enumerate}
\vspace{2\lineskip}

There is no shortage of literature on quantum graphs: a more
in-depth introduction to quantum graphs with a focus on universal
spectral statistics can be found in the review \cite{GnuSmi_ap06};
several reviews \cite{AkkComDes_ap00,Kuc_incol08,andrade2016green,BolKer_manybodyQG_2020}
cover other physical and mathematical aspects. There is also a number of textbooks that
address various aspects of spectral theory of quantum graphs
\cite{BerKuc_graphs,post2012spectral,ExnerKovarik_QuantumWaveguides2015,Mugnolo_book,Kurasov_SpectralGeometryBook},
as well as collections of papers \cite{BerCarFulKuc_eds06,ExnKeaKuc_eds08}.
Finally, there is a number of pedagogical introductory surveys
\cite{Kuc_wrm04,Kuc_jpa05,Ber_gcst17,band2018quantum}.

Apart from their role in quantum chaos and spectral theory,
quantum graph models have many applications in science beyond quantum
theory, starting with quantum chemistry
\cite{Pau_jcp36,RueSch_jcp53,Gri_tfs53,AmoLeyMar_jmc04}, to physics of
conductivity \cite{ExnSeb_incol89,Rub_incol06,GolGas_prb08}, to
photonic crystals \cite{FigKuc_siamjam98,KucKun_em99,Kuc_incol01,KucKun_acm02}.
Experimentally, wave equations on a network may be realised with microwave
cables \cite{hul2004experimental}, optical fibres
\cite{babaee2015wave}, elastic strings, and other types of waveguides.
Similar techniques contribute to studies of other differential
operators, in the modelling of beam frames
\cite{Lagnese_multilink,brewer2018elastodynamics,BerEtt_sam22}, blood
flow \cite{Carl_incol06}, chemotaxis \cite{SheSheSuk_pafa24}, and
spread of epidemics \cite{KraDurBri_bmb24}.  An overview of some
generalizations of quantum graph models using more general linear and
nonlinear wave equations can be found in \cite{kairzhan2022standing}.
Finally, much work has been done on mathematical justification of the
use of quantum graphs as limits for thin waveguides
\cite{RueSch_jcp53,post2012spectral,Sai_jde00,KucZen_jmaa01,KucZen_incol03,Grieser_LMS,Gri_incol08,ExnPos_jpa09,ExnPos_cmp13},
with particular attention paid to the types of vertex conditions one
obtains in the limit.

\emph{We do not aim to give a complete overview of all current research that
uses quantum graph models or their generalizations. Rather we use a
selection of topics that we believe to be suitable as a starting point
to explore the field, especially from a quantum chaos perspective.}
% While many more quantum graph topics could have been mentioned we
% wanted to keep the introduction relatively short.  Still, we have
% selected a few topics beyond quantum chaos according to our own
% interests and expertise that we believe to be interesting in a first
% introduction to quantum graphs.
Accordingly our list of references is a selection to work that is
either relevant to the selected topics or contain further literature
that explore the topics in a more complete way.

The manuscript is organised as follows. In \Cref{sec:basic_spectral_theory}
we define quantum graphs and give a didactical introduction to the spectral theory of
quantum graphs with an emphasis on concepts that are relevant in quantum chaos
such as the scattering approach, the trace formula and expansions into trajectories and
periodic orbits.
An overview of some applications to quantum chaos are
given in   \Cref{sec:spectrum} (spectral statistics and universal 
random-matrix behaviour)
 and \Cref{sec:wavefunctions} (Quantum ergodicty, topological resonances and nodal statistics).
In \Cref{sec:applications} we (very briefly) 
introduce the quantum graph approach to Fourier quasicrystals
and metamaterials as two interesting 
current directions (one mathematical, the other very applied). 
Finally, in chapter \Cref{sec:mathematical-applications}
we give another didactical introduction to
 mathematical techniques that complement our first introduction
 in \Cref{sec:basic_spectral_theory}.

\section{Quantum graphs, their stationary states and energy spectrum}
\label{sec:basic_spectral_theory}

\subsection{A quantum graph as a metric graph with a Schr\"odinger
  operator}
\label{sec:metric_graph}

A \emph{metric graph} $\Graph$ is described most succinctly as a set
$\Eset$ of open intervals $\me_j := (0,\ell_j)$ --- the edges of the
graph --- connected together at vertices $\mv\in\Vset$.  The length
$\ell_j\equiv \ell_{\me_j}$ of the edge $\me_j$ is positive or infinite; edges of
infinite length are called \term{leads} and those of finite length \term{bonds}.  
Each vertex $\mv$ is
understood as a set (or equivalence class) of endpoints of the
edges.  On each edge \(\me\in \Eset \), we have a local coordinate
\(x_{\me} \in (0,\ell_{\me})\), with the choice of direction arbitrary, but
fixed.  The two endpoints \(x_{\me}=0\) and \(x_{\me}= \ell_{\me}\) are referred
to as \textit{origin} \({o}(\me)\) and \textit{terminus} \({t}(\me)\).
Standard assumptions in the field and used in the remainder of this manuscript are that

\vspace{2\lineskip}
\begin{itemize}
\item there is a finite or countable number of edges,
\item each endpoint of each edge belongs to one and only one vertex
  and
\item each vertex contains at least one endpoint.
\end{itemize}
\vspace{2\lineskip}

\noindent
Mathematically, the above construction is often expressed by saying
that vertices are equivalence classes of endpoints of edges.
The orientation of edges, implicitly assumed in the interval notation,
will not affect any of the quantities we discuss here.

One now considers the Hilbert space of functions defined on the edges
(intervals) $\me$
\[
  L^2(\Graph):=\bigoplus_{\me\in\Eset} L^2(\me)
  = \bigoplus_{\me\in\Eset} L^2 \big((0, \ell_\me)\big),
\]
with the natural inner product.  On this Hilbert space we consider the
Schr\"odinger Hamiltonian, the differential operator of the form
\begin{equation}
  \label{eq:schrod_def}
  H: \psi = \left(\psi_1, \ldots, \psi_{|\Eset|}\right)
  \mapsto
  \left( \left(-\frac{d^2}{dx^2} + V_1\right) \psi_1, \ldots,
    \left(-\frac{d^2}{dx^2} + V_{|\Eset|}\right)\psi_{|\Eset|} \right).
\end{equation}
Here $V_1,\ldots,V_{|\Eset|}$ are sufficiently nice functions on the
edges of the graph, describing the potential along the edges.  We will
usually assume $V\equiv 0$, effectively restricting our attention to
the Laplace operator
\begin{equation}
  \label{eq:laplace_def}
  H: \psi = \left(\psi_1, \ldots, \psi_{|\Eset|}\right)
  \mapsto
  \left(-\frac{d^2}{dx^2}\psi_1, \ldots,
    -\frac{d^2}{dx^2}\psi_{|\Eset|}\right).
\end{equation}
In order to give meaning to the time-dependent 
Schr\"odinger equation \( i \frac{\partial}{\partial t} \Psi(t)= H \Psi(t)\)
one needs to define \(H\) carefully and introduce matching conditions at the vertices
that ensure quantum probability conservation -- mathematically this means that
\(H\) needs to be self-adjoint. Once we have defined self-adjoint Hamiltonian we
will focus on stationary solutions \(H\psi(x) = E \psi(x)\) where \(E\) is in the
spectrum of (generalized) eigenvalues.

%\SG{In the following paragraph you refer to operators $H$ and $L$ -- I believe you 
%wanted to call the Laplacian $L$. Since $L$ is already used for lengths and  in $L^2(\Gamma)$. 
%So I think keeping $H$ for the Laplacian is fine. In any case the definition of the is the
% same with or without nice potentials. I have commented out the references to $L$
%} \GB{Sorry, I came to the conclusion that we should keep the operator as $H$ for
 % Hamiltonian, whether it has a potential or not --- but I didn't do a
 % good job implementing this.  Thanks!}
Not all functions in $L^2(\Graph)$ can be differentiated, so the
\emph{domain} of the operator %s $L$ and
$H$ is restricted to those functions whose second derivative is again
in $L^2(\Graph)$, namely the Sobolev space $\bigoplus_{\me}H^2(\me)$.
It is further restricted by \emph{vertex conditions} to make the
operator %s $L$ and
$H$ \emph{self-adjoint}.  There is a significant mathematical
literature dedicated to classifying all possible self-adjoint
conditions (see, for example,
\cite{Rof_tffap69,KosSch_jpa99,Har_jpa00,Kuc_wrm04} for the full
classification and the textbooks \cite{BerKuc_graphs,
  Kurasov_SpectralGeometryBook} for further references); we will
briefly summarize one parametrization in Section~\ref{sec:selfadjoint}
below.  However, we will restrict ourselves to the so-called
\emph{$\delta$-type conditions} defined by two
requirements at each vertex $\mv$, which we assume to be the 0 point
of the edges $\me_{j_1},\ldots,\me_{j_m}$ and the final point $\ell$
of the edges $\me_{j_1'},\ldots, \me_{j'_n}$:

\vspace{2\lineskip}
\begin{enumerate}
\item \emph{Continuity of the wave function:} \\
  The value of the wave function is continuous at the vertex: every
  $\psi_j$ for every edge $\me_j$ incident to $\mv$ takes a single,
  well-defined value $\psi(\mv)$ at the vertex. 
  \begin{equation}
    \label{eq:cond_continuity}
    \psi_{j_1}(0) = \dots = \psi_{j_m}(0)
    = \psi_{j_1'}(\ell_{j_1'}) = \ldots = \psi_{j_n'}(\ell_{j_n'}) =: \psi(v).
  \end{equation}
    \item \emph{Derivative jump:} \\
    The sum of the outgoing derivatives of the wave function at the 
    vertex is proportional to the value of the function at that vertex.
    \begin{equation}
      \label{eq:cond_current}
      \frac{d}{dx}\psi_{j_1}(0) + \dots + \frac{d}{dx}\psi_{j_m}(0)
      -\frac{d}{dx}\psi_{j_1'}(\ell_{j_1'}) - \ldots -
      \frac{d}{dx}\psi_{j_n'}(\ell_{j_n'})
      = \alpha_\mv \psi(v)      
    \end{equation}
    Here, $\alpha \in \R$ is a real \emph{coupling constant}
    or the \emph{strength of the $\delta$-potential}.  Note that the
    signs in \eqref{eq:cond_current} are arranged so that the 
    derivative is taken in the direction away from the vertex and into the edge.
\end{enumerate}
\vspace{2\lineskip}

\noindent
We would like to highlight two important special cases:

\vspace{2\lineskip}
\begin{itemize}
\setlength{\leftmargin}{0.1\textwidth}
\item \emph{Neumann--Kirchhoff Conditions ($\alpha_\mv=0$):} This is
  the most common condition, often called ``standard''.  It
  is analogous to Kirchhoff's current law in electrical circuits.
\item \emph{Dirichlet Conditions ($\alpha \to \infty$):} \\
  The limit can be taken by first dividing equation~\eqref{eq:cond_current}
  by $\alpha_\mv$; it leads to the Dirichlet condition:
  \begin{equation}
    \label{eq:cond_Dirichlet}
    \psi(\mv) = 0, 
  \end{equation}
  Note that this condition effectively disconnects the edges at the
  vertex, thus altering the topology of the graph.
\end{itemize}
\vspace{2\lineskip}

The energy form of $H$ (for the
Laplacian, or Schr\"odinger operators with bounded potentials) is
\begin{equation}
  \label{eq:energy_form}
  \mathfrak{h}_{\alpha}[\psi]
  = \sum_{\me\in\Eset}\int_{\me}\Big(|\psi_\me'|^2+V_\me|\psi_\me|^2\Big)\,dx
  + \sum_{\mv\in\Vset}\alpha_{\mv}\,|\psi(\mv)|^2,
\end{equation}
with domain consisting of functions $\psi\in\bigoplus_{\me}H^1(\me)$
that are continuous at all vertices.  
For a normalized function $\|\psi\|=1$ 
the energy form is just the energy expectation value written
in a symmetric way using integration by parts.
Through the use of
Courant--Fischer minimax, the energy form can provide estimates on the
eigenvalues of $H$, see \Cref{sec:spec_estimates}.  The eigenfunctions
are the critical points of $\mathfrak{h}_{\alpha}[\psi]$ on the unit
sphere $\|\psi\|=1$ and the eigenvalues are the corresponding critical
values.

One observation we can make from \eqref{eq:energy_form} is that the
second term may be incorporated in the first term by adding
$\alpha_\mv \delta(x-\mv)$ to the potential $V_\me$ (for one of the
edges $\me$ incident to the vertex $\mv$).  This observation can be
made mathematically rigorous
\cite{Exn_jpa96,AlbeverioKurasov_singpert} and is the reason why,
after \cite{Exn_jpa96}, the vertex conditions
\eqref{eq:cond_continuity}-\eqref{eq:cond_current} are known as the
\term{$\delta$-type conditions}.

%\begin{example}
%  \label{ex:3star}
%  Consider the star graph with $|\Eset|=3$ edges, which we represent as
%  the intervals $(0,\ell_j)$ identified at the $\ell$-endpoint, where
%  $\delta$-type condition is imposed.  At the endpoint $0$ on every
%  edge we impose the Dirichlet conditions.  To summarize,
 % \begin{align}
%    \label{eq:star_conditions}
%    &\psi_j(0) = 0, \quad j\in\{1,\ldots,|\Eset|\}, \\
%    &\psi_1(\ell_1) = \psi_2(\ell_2) = \psi_3(\ell_3) \equiv \Psi_0, \\
%    &\sum_{j=1}^{3} \left(-\frac{d\psi_j}{dx_j}\right)\!\bigg|_{x_j=\ell_j} = \alpha \Psi_0
%  \end{align}
%\end{example}
%\GB{I don't think this example illustrates anything important here...
%  I would move it later where we can talk about the spectrum}

\subsection{Scattering approach to stationary states and the spectrum of quantum graphs}

Let us consider from now on a quantum graph with vanishing potentials
\(V_j=0\) on all edges such that the Schr\"odinger Hamiltonian just reduces
to the Laplacian. 
The spectrum and the corresponding stationary states are then described by
the free Schr\"odinger equation 
\(
  -\frac{d^2}{dx^2}\psi(x) = E \psi(x)
\) 
on each edge and matching conditions  (as defined above in \Cref{sec:metric_graph})
at the vertices.
In the scattering approach one focuses on the positive part of the spectrum
\(E=k^2>0\) where \(k>0\) is the wave number. 
It is based on the following two observations:

\vspace{2\lineskip}
%\vskip

%\begin{tcolorbox}
\begin{enumerate} 
\item The stationary states of are locally described by 
superpositions of two plane waves 
\(\psi(x)= a e^{ikx}+b e^{-ikx}\)
on each edge of the graph.
\item At the vertices the plane waves scatter to other edges with amplitudes 
that can be derived from the matching conditions.
\end{enumerate}

\vspace{2\lineskip}

We will develop the scattering approach  in some detail here 
for Neumann-Kirchhoff conditions (as defined above in Sec~\ref{sec:metric_graph})
at each vertex.
We will focus on describing the spectrum and the corresponding stationary
states. The approach may also be used to describe Green functions (the kernel
of the resolvent of the Hamilton operator) -- for this we refer to the literature \cite{BarGas_pre02,lawrie2023closed}.

Before we move forward it is useful to introduce some additional notation.
We say that an edge \(\me\in \Eset\) is \textit{adjacent}
to a vertex \(v\in \Vset\) if the latter is one
of the endpoints. 
Two edges \(\me\) and \(\me'\) are adjacent
if they share
a common endpoint vertex and two vertices are adjacent
if there is an edge that is adjacent to both vertices.
The \emph{star} \(\mathsf{S}_v\) at vertex \(v\) is the set of all edges adjacent to
\(v\).
The \emph{degree} \(d_v\) of a vertex
is the number of endpoints of all adjacent edges that correspond to the vertex \(v\).
A \emph{loop} is a bond for which the two endpoints are at 
the same vertex \(v\) 
(each loop counts twice towards the degree of the adjacent vertex).
A \emph{dangling bond} has at least one endpoint with degree one. 
We only consider \textit{connected} graphs -- graphs that are not a union of two or more 
graphs with no adjacent edges among them.
The set of bonds is denoted by \(\Bset\) and the set of leads by \(\Lset\).
The set of all edges is then a disjoint union
\(\Eset=\Lset \cup \Bset \)
and the number of edges is
 \(\lvert \Eset\rvert=\lvert \Bset\rvert + \lvert \Lset \rvert \). 
 
In this chapter we will consider two types of connected graphs 

\vspace{2\lineskip}
\begin{enumerate}
\item
\textit{Closed (compact) quantum graphs} have a finite number of bonds and no leads,
So \(\lvert \Lset\rvert=0\) and \(\lvert \Eset \rvert=\lvert \Bset \rvert\). 
\item
\textit{Scattering graphs (open) quantum  graphs} also have a finite number 
of bonds \(\lvert \Bset\rvert \ge 0 \) but have at least one edge
\(\lvert \Lset\rvert \ge 1  \).
\end{enumerate}
\vspace{2\lineskip}

\noindent
There are interesting graph structures beyond these two classes, for instance
\textit{periodic graphs}. These have no leads and a countable infinite number of bonds 
that are arranged in a repeating 
pattern of a lattice. One may add disorder to periodic graphs by choosing the edge lengths
(or matching conditions) independently from some ensemble which of course breaks the 
strict periodicity. 

For each edge \(\me\), a plane wave can travel in two possible
directions, necessitating the following notation: \(\me_+\) will refer
to the \term{directed edge} with the same orientation as $\me$ and
\(\me_-\) will refer to the \term{directed edge} with to orientation
opposite to $\me$.  We refer to the set of directed edges as
\(\Eset_{\pm} \).  The origin and terminus of a directed edges are
defined as \({o}(\me_+)=t(\me_-)= o(\me)\) and \({o}(\me_-)=t(\me_+)=
t(\me)\).  We occasionally use $\me_{j,+}$ and $\me_{j,-}$ for the two
directed edges corresponding to $\me_j \in \Eset$.
We say that \(\me_{j,s}\) \term{follows} \(\me_{\theta,\sigma}\) (and
write \(\me_{\theta,\sigma} \to \me_{j,s}\)) if
\(o\left(\me_{j,s}\right)=t\left(\me_{\theta,\sigma}\right)\).

Coming back to the superposition of two plane waves on a given edge, let us consider 
a bond \(\me \in \Bset\) and write  
\begin{equation}	
	\begin{split}
	\psi_{\me}(x_{\me})=& a^{\mathrm{in}}_{\me_-}e^{-ikx_{\me}}+
	a^{\mathrm{out}}_{\me_+}
    e^{ikx_{\me}}\\
    =&
    a^{\mathrm{out}}_{\me_-}e^{ik(\ell_{\me}-x_{\me})}+
    a^{\mathrm{in}}_{\me_+}
    e^{-ik(\ell_{\me}-x_{\me})} .
    \end{split}
\end{equation}
Here \(a^{\mathrm{in}}_{\me_-}\) and  \(a^{\mathrm{out}}_{\me_+}\)
is the incoming and outgoing amplitudes  at \(o(\me)\). At the other endpoint
\(t(\me)\) one then has the incoming and outgoing amplitudes
\begin{equation}
   a^{\mathrm{in}}_{\me_+}= a^{\mathrm{out}}_{\me_+}  
   e^{ik\ell_{\me}}, \quad \text{and} \quad
   a^{\mathrm{out}}_{\me_-}= a^{\mathrm{in}}_{\me_-}e^{-ik\ell_{\me}}\ .
\end{equation}
On any bond the outgoing directed edge from one endpoint is incoming
at the other end.  The plane waves at these two endpoints are just
related by a phase factor.  Here, \(a^{\mathrm{in/out}}_{\me_\pm}\) is
the complex wave amplitude on edge $\me$ propagating in the direction
of increasing (\(+\)) or decreasing (\(-\)) position \(x_{\me}\),
heading \(\mathrm{in}\) or \(\mathrm{out}\) of a vertex. If \(\me\) is
a lead only
the amplitudes \(a^{\mathrm{in/out}}_{\me_\pm}\) at \(x_{\me}=0\) are used.\\
Introducing the $\Eset_\pm$-indexed diagonal length matrix
\(\mathbf{L}\) with matrix elements
\begin{equation}
  \label{eq:Ldef}
  \mathbf{L}_{\me_{j,s}, \me_{j,s}} = \ell_j,
\end{equation}
the in- and outgoing 
wave amplitudes can be mapped to one another 
by the diagonal 
square \(2\lvert \Bset \rvert\)-dimensional
matrix 
\begin{equation}
  \label{eq:Tdef}
  \mathbf{T}(k)=e^{i k\mathbf{L}}
\end{equation}
that takes account of the phase difference 
between wave 
amplitudes:
\begin{equation}
    \mathbf{a}^{\mathrm{in}}_{\Bset}= \mathbf{T}(k)\  
    \mathbf{a}^{\mathrm{out}}_{\Bset}\ .
    \label{bond_transport}
\end{equation}
Here, \(\mathbf{a}^{\mathrm{in/out}}_{\Bset}\) refers to the 
\(\lvert \Bset\rvert\)-dimensional column vector
of plane wave coefficients on the directed bonds. \\
In addition, the graph wave amplitudes can be mapped onto one another 
across the 
vertices by taking account of the imposed vertex boundary conditions.
This is the starting point of the construction and we now need add
the matching conditions at the vertices to determine the spectrum and the corresponding
stationary wave functions.

\subsubsection{The vertex scattering matrix}

Let us first consider a single vertex
\(v\) of degree \(d_v\). Combining the corresponding \(d_v\) incoming and outgoing
amplitudes in column vectors we claim that the Neumann-Kirchhoff conditions are equivalent 
to the condition
\begin{equation}
\mathbf{a}^{(v),\mathrm{out}}= \boldsymbol{\sigma}^{(v)} 
\mathbf{a}^{(v),\mathrm{in}}
\end{equation}
in terms of the  \(d_v \times d_v\)
\textit{vertex scattering 
matrix}
\begin{equation}\label{Kirchoff-Neumann BC}
    \boldsymbol{\sigma}^{(v)}=  - \mathbb{I}_{d_v} +
    \frac{2}{d_v}\mathbb{E}_{d_v},
\end{equation}
where \(\mathbb{I}_{d_v}\) is the identity matrix and \(\mathbb{E}_{d_v}\) is the 
matrix 
of dimension \(d_v\) with all entries equal to one. 
To show this we may consider without loss 
of generality the star \(\mathsf{S}_v\) with \(d_v\) leads 
such that \(x_e=0\) at the vertex for all adjacent edges. 
For the general case where \(\mathsf{S}_v\) contains bonds (and possibly loops) one just has to
observe that we only consider the wave function and its derivative at the vertex.
Given an arbitrary set of incoming amplitudes the
wave function on the adjacent edges are
\begin{equation}
\begin{split}
\psi_{\me}(x_{\me})=& a_{\me_-}^{\mathrm{in}} e^{- i k x_{\me}}
+ a_{\me_+}^{\mathrm{out}} e^{i k x_{\me}}
\\
=&a_{\me_-}^{\mathrm{in}} \left(e^{- i k x_{\me}}-e^{ikx_{\me}}\right)
+\frac{2}{d_v} \sum_{\me'\in \mathsf{S}_v } a_{\me'_-}^{\mathrm{in}} e^{i k x_{\me}}\ .
\end{split}
\end{equation}
This indeed satisfies continuity as
\begin{equation}
	\psi_{\me}(0)=\frac{2}{d_v} \sum_{\me'\in \mathsf{S}_v } a_{\me'_-}^{\mathrm{in}}
\end{equation}
is independent of \(\me\). It also satisfies the 
derivative condition
\begin{equation}
  \frac{1}{ik} \sum_{\me \in \mathsf{S}_v} \frac{d \psi_{\me}}{dx_{\me}} (0)=
  -2 \sum_{\me \in \mathsf{S}_v} a_{\me_-}^{\mathrm{in}} +2 
  \sum_{\me' \in \mathsf{S}_v} a_{\me'_-}^{\mathrm{in}}=0\ . 
\end{equation}

One may combine all vertex scattering matrices into a single 
(directed) edge scattering matrix \(\mathbf{S}\), such that
\begin{equation}
  \mathbf{a}^{\mathrm{out}}=  \mathbf{S} \; 
  \mathbf{a}^{\mathrm{in}}.
  \label{graph_scattering}
\end{equation}
Here, \(\mathbf{a}^{\mathrm{in/out}}\) is a \(2 \lvert \Bset\rvert + \lvert\Lset \rvert\) 
dimensional column vector 
of all the incoming/outgoing 
amplitudes for all bonds and leads.
The scattering matrix elements are expressed in terms of the individual 
vertex 
scattering matrices \(\boldsymbol{\sigma}^{(v)} \), such that
\begin{equation}
\label{scattering_matrix}
\mathbf{S}_{\me_s \me'_{s'}} = 
\begin{cases}
   \sigma^{(v)}_{\me \me'}
   & \text{if \(\me'_{s'} \to \me_s\) and \(v=t(\me'_{s'})=o(\me_s)\),}\\
   0 & \text{else.}
\end{cases}
\end{equation}
Note that the non-zero elements of \(\mathbf{S}\)
fully describe the connectivity of the graph.
Equations \eqref{bond_transport} and \eqref{graph_scattering} 
reduce the problem of finding the energy eigenstates 
to a set of linear equations.

\subsubsection{The discrete spectrum and corresponding stationary states for 
closed quantum graphs}

In the case of a closed quantum graph, we have 
\(\mathbf{a}^{\mathrm{in/out}}_{\Bset} \equiv 
\mathbf{a}^{\mathrm{in/out}} 
\). 
In this case 
the two relations \eqref{bond_transport} and \eqref{graph_scattering}  
combine 
to give one condition, 
\begin{equation}
    \mathbf{a}^{\mathrm{in}} = \mathbf{U}(k)\  \mathbf{a}^{\mathrm{in}}, 
    \label{quantization}
\end{equation}
with the \(2\lvert \Bset \rvert\) dimensional \textit{quantum map}
\begin{equation}
    \mathbf{U}(k)= \mathbf{T}(k) \mathbf{S}\ .
    \label{quantum_map}
\end{equation}
Non-trivial solutions 
to \eqref{quantization} exist for wave numbers \(k\) for which the quantum 
map 
\(\mathbf{U}(k)\) has a 
unit eigenvalue, and we arrive to the following criterion \cite{KotSmi_prl97,KotSmi_ap99,KosSch_jpa99,Bel_laa85,BolEnd_incol08,BolEnd_ahp09}:\\
  The positive (discrete) energy spectrum \(E=k^2>0\) of the quantum
  graph is given, together with multiplicity, by the
  positive zeros $k$ of the \(\xi(k)\) \term{secular equation}
  \begin{equation}
    \label{secular}
    \xi(k) :=
    \det\left(\mathbb{I}_{2\left| \Bset \right|}-\mathbf{U}(k)\right)
    = \det\left(\mathbb{I}_{2\left| \Bset \right|}
      -\mathbf{T}(k) \mathbf{S} \right)
    = 0\ .
  \end{equation}
  The function \( \xi(k)\) is known as the \textit{secular function}.
  Additionally, the positive energy $E=k^2$ eigenstates of the free
  Schr\"odinger Hamiltonian $H = -\frac{d^2}{dx^2}$ can be obtained
  from \eqref{quantization} which has as many linearly independent
  non-trivial solutions \(\mathbf{a}^\mathrm{in}\) as the multiplicity
  of $E$.

The secular equation is the basis for the derivation of the trace
formula, Section~\ref{sec:trace_formula} and for the Barra--Gaspard
approach to eigenvalue and eigenvector statistics, Section~\ref{sec:BG}.
While we derived the secular equations for quantum graph with
Kirchhoff-Neumann conditions, a similar equation holds in more general
circumstances.  The presence of Dirichlet conditions alter the matrix
$\mathbf{S}$: the Dirichlet vertex scattering matrix is
$\sigma^{(\mv)} = (-1)$, in contrast to $(+1)$ predicted by
\eqref{Kirchoff-Neumann BC} for the Neumann vertex).  Importantly,
$\mathbf{S}$ remains constant in this case.  More general
$\delta$-type conditions introduce $k$-dependence of the matrix
$\mathbf{S}$, making the analytical tools much more difficult to
apply.

The secular equation typically gives incorrect answers for the energy
$E=0$; this discrepancy was analysed in
\cite{FulKucWil_jpa07,Kur_ark08}.  For connected Neumann--Kirchhoff
graphs, the constant function \(\psi_{\me}(x_{\me})=1\) is the unique
eigenstate of the Hamiltonian with energy \(E=k^2=0\).

The second observation is that for graphs with non-negative
$\delta$-type conditions ($\alpha_\mv \geq0$), including
Neumann--Kirchhoff and Dirichlet, the energy expectation
$\mathfrak{h}_{\alpha}[\phi]$ from \eqref{eq:energy_form} is
non-negative (note that $V \equiv 0$), showing that there are no
negative energy eigenvalues.
% \begin{equation}	
% -\sum_{e \in \Bset} \int_{0}^{\ell_e}\phi_{e}(x_e)^*  \frac{d^2 \phi_{e}}{dx^2}(x_e) dx_e
% = 	 \sum_{e \in \Bset} \int_{0}^{\infty} \left|\frac{d \phi_{e}}{dx}(x_e)\right|^2 dx_e \ge 0
% \end{equation}
% This shows that there are no negative energy eigenvalues and that
% the constant function with  \(E=0\) 
% is the non-degenerate ground state.

\begin{figure}[t]
  \centering
  \begin{tikzpicture}[line cap=round, line join=round, scale=1]
    % --- vertices ---
    \coordinate (v) at (0,0);        % the unique vertex (loop endpoints identified here)
    \coordinate (u) at (-3.2,0);     % degree-1 endpoint of e1

    % draw vertices
    \fill (v) circle (2.2pt) node[below] {$\mv_2$};
    \fill (u) circle (2.2pt) node[below] {$\mv_1$};
    
    % --- edge e1 (a bond/interval) ---
    \draw[thick] (u) -- (v);
    \node[above] at ($(u)!0.5!(v)$) {$\me_1$};
    
    % --- edge e2 (a loop attached at v) ---
    % draw a loop based at v
    \draw[thick] (v) .. controls (2,2.4) and (2,-2.4) .. (v);
    \node at (1.75,0) {$\me_2$};
  \end{tikzpicture}
  \hspace{2cm}
  \begin{tikzpicture}[scale=1.0, line cap=round, line join=round]
    % Parameters
    \def\R{{2.6, 2.3, 1.9, 2.6}}      % edge length (visual)
    \def\n{3}        % number of edges (illustration of |\mathcal E|)
    \def\vdot{2.2pt} % vertex dot size
    
    % Central (identified) vertex at x=\ell_j
    \coordinate (V) at (0,0);
    \fill (V) circle (\vdot);
    \node[above right=2pt] at (V) {$v$};

    % Draw edges as intervals (0,\ell_j) with 0 at the outer endpoints
    \foreach \k in {1,...,\n} {
      \pgfmathsetmacro{\ang}{90 + (\k-1)*360/\n} % spread edges
      \coordinate (Pk) at (\ang:\R[\k]);    
      % edge
      \draw[thick] (V) -- (Pk);
      % outer endpoint corresponding to x=0
      \fill (Pk) circle (1.4pt);
      \node at (\ang:\R[\k]+0.35) {$\ell_\k$};
    }
    % label near the edge (interval + length endpoint)
    \node[right] at (90:\R[1]*0.62) {$\me_{1}$};
    \node[above] at (210:\R[2]*0.62) {$\me_{2}$};
    \node[above] at (330:\R[3]*0.62) {$\me_{3}$};
  \end{tikzpicture}
  \caption{
  Left: A
    tadpole graph from Example~\ref{ex:tadpole}. 
  Right:
  A star graph with three bonds, illustrating Example~\ref{ex:star} with \(N=3\).}
  \label{fig:star}
\end{figure}
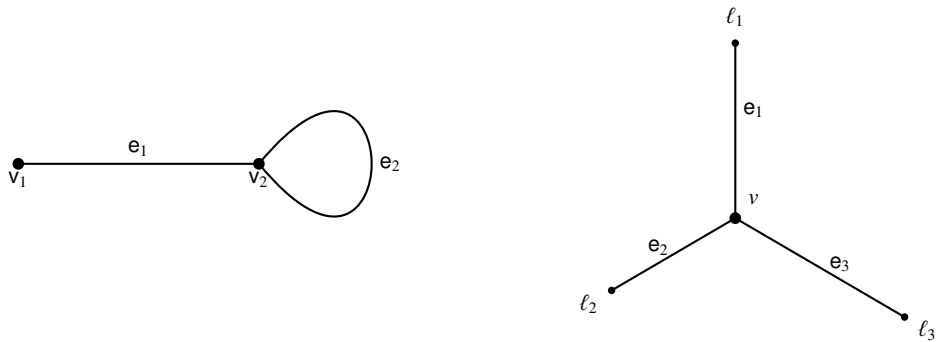

\begin{example}
  \label{ex:tadpole}
	The \textit{quantum tadpole graph} (also known as quantum lasso graph) 
	consists of two bonds
	\(\me_1\) and \(\me_2\)
	of length \(\ell_1\) and \(\ell_2\), see Fig.~\ref{fig:star}(left). 
	The edge \(\me_2\) is a loop and thus starts and ends at the same vertex.
	The edge \(\me_1\) is a dangling bond. At one end (which we take to be the origin) 
	it is connected to the  loop \(\me_2\), such that 
	\(o(\me_1)=o(\me_2)=t(\me_2)\).
	The other end \(t(\me_1)\) is a vertex of degree 1.
	Let us order the directed bonds by writing 
	\begin{equation}
	\mathbf{a}^{\mathrm{in}}= 
	\begin{pmatrix}
	a^{\mathrm{in}}_{\me_{1+}}\\
	a^{\mathrm{in}}_{\me_{2+}}\\
	a^{\mathrm{in}}_{\me_{1-}}\\
	a^{\mathrm{in}}_{\me_{2-}} 
	\end{pmatrix}\ .
	\end{equation}
	The matching conditions and the connectivity of the graph then lead to
	\begin{equation}
	\mathbf{S}=
	\begin{pmatrix}
	0&\frac{2}{3}&-\frac{1}{3}&\frac{2}{3}\\
	0&\frac{2}{3}&\frac{2}{3}& -\frac{1}{3}\\
	1&0 & 0 & 0\\
	0&-\frac{1}{3}&\frac{2}{3} & \frac{2}{3}
	\end{pmatrix}
	\end{equation}
	and the secular function
	\begin{equation}
	\xi(k)= \frac{1}{3}\left(1- e^{ik \ell_2} \right)\left(3-
	e^{ik\ell_2}+e^{i2k\ell_1}-3e^{ik(2\ell_1+\ell_2)} \right)\ .
	\end{equation}
	As the secular function factorises, the spectrum can be decomposed into two parts.
	In the first part we have wave numbers \(k= \frac{2\pi}{\ell_2}n\) for positive integers \(n\).
	In the second part we have the (positive) wave numbers that are implicitly given as the
	solutions of 
	\begin{equation}
          \label{eq:sec_tadpole_sym}
          \sin\left(\frac{2\ell_1-\ell_2}{2} k\right)=
          3\sin\left(\frac{2\ell_1+\ell_2}{2} k\right)\ .
	\end{equation}
	For \(k= \frac{2\pi}{\ell_2}n\) the corresponding energy
        eigenstates are of the form \(\psi_1(x_1)=0\) and
        \(\psi_2(x_2)= \sqrt{\frac{2}{\ell_2}}\sin\left(\frac{2\pi n
            x_2}{\ell_2}\right) \).  They vanish identically on edge
        \(e_1\) (including the two endpoint vertices) and are
        supported on the loop \(e_2\).  In general energy eigenstates
        that are supported on a proper subgraph are called
        \textit{perfect scars}. The existence of perfect scars on
        subgraphs is a special property of quantum graphs and we will
        come back to them in \Cref{sec:topological_resonances}. In larger quantum graphs
        they may exist on subgraphs that consist of more than one edge
        -- loops are just the most simple example where
        Neumann-Kirchhoff conditions always lead to perfect scars
        (this generalises to the \(\delta\)-type conditions discussed
        above).  The energy eigenstates that correspond to the second
        part with the implicit condition
	are generally supported on the whole tadpole graph.\\
	An alternative way to obtain the partition of the full
        spectrum into the two parts described above is by
        symmetry. Indeed the graph has a `mirror' symmetry along an
        axis along the bond \(\me_1\) and the midpoint of the bond
        \(\me_2\); in Figure~\ref{fig:star}(left) this corresponds to
        vertical reflection.  This implies that eigenstates are
        either odd or even under the symmetry operation -- and the
        corresponding spectra are just the ones described above (the
        perfect scars on the loop are odd).  For graphs with richer
        symmetry groups, an expansion into spectra corresponding to
        all irreducible representation has been developed in
        \cite{BanParBen_jpa09,ParBan_jga10} and
        \cite[Sec.~4]{BanBerJoyLiu_prep17}.
\end{example}

\begin{example}
\label{ex:star}
    Next let us consider a quantum star graph with 
    \(N\equiv \left|\Bset\right|=\left|\Eset\right|\)  bonds \(\me_1,\me_2,\dots, \me_N\)
    of lengths \(\ell_k\), 
    see  Fig.~\ref{fig:star}(right) for the case \(N=3\). 
    All bonds are dangling bonds and have one endpoint 
    vertex in common that we choose to be 
    the origin \(o(\me_1)=o(\me_2)=\dots=o(\me_N)\). 
    The scattering at all vertices is described as
    \begin{equation}
    \mathbf{S}=
    \begin{pmatrix}
    		0_N & -\mathbb{I}_N + \frac{2}{N} \mathbb{E}_N\\
    		\mathbb{I}_N & 0_N
    \end{pmatrix}
    \end{equation}
    (where \(0_N\) is the zero matrix of dimension \(N\times N\))
    and we get the secular equation
    \begin{equation}
    \xi(k)=
    \mathrm{det}\left(\mathbb{I}_{2N}- \mathbf{T}(k) \mathbf{S}\right)
    =\mathrm{det}\left(\mathbb{I}_N - \mathbf{T}_{\mathrm{red}}^2(k) 
    \left(-\mathbb{I}_N + \frac{2}{N} \mathbb{E}_N\right)
    \right)=-\frac{i 2^N e^{i k L_{\Graph}}}{N}
    \left[\prod_{n=1}^N \cos(k \ell_n) \right]
    \left[\sum_{n=1}^N \tan(k \ell_n) \right]
    \end{equation}
    where \(\mathbf{T_{\mathrm{red}}(k)}= 
    \mathrm{diag}\left(e^{ik\ell_1},\dots, e^{ik\ell_N} \right)\).
    Here \(L_{\Graph}= \sum_{n=1}^N \ell_n\) is the total length of the graph.
    The wave number spectrum of a star graph
    is determined by the (positive) zeros of \(\xi(k)\).  
    Note that the factor \( \prod_{n=1}^N \cos(k \ell_n)\)
    has zeros at the same locations as the poles of the factor
    \(\sum_{n=1}^N \tan(k \ell_n) \).  Whereas the poles are always
    simple, the zeros may have higher order --- and then the product
    is zero. This implies that the spectrum consists of two  
    parts. The first part consists of the positive solutions of \(\sum_{n=1}^N \tan(k \ell_n)=0 \)
    (with exactly one solution between two neighbouring poles). The second part
    consists of any \textit{degenerate} solutions of \( \prod_{n=1}^N \cos(k \ell_n)=0\)
    that is there should be 
    \(s\ge 2\) cosine factor that vanish simultaneously.	 The degeneracy of the 
    corresponding wave number in the spectrum is then \(s-1\).
    Note that in this case there is a nodal point on the central vertex and the 
    corresponding eigenstate may be supported on a subset of the bonds.
    If all bond lengths are rationally independent (that is one cannot write 
    \(\sum_{n=1}^N m_n \ell_n=0\) for any non-trivial integers \(m_n\)) then
    the second part of the spectrum is empty and no eigenstate has a nodal point on
    the central vertex.\\
    Let us conclude this chapter by discussing the special case when all bond
    lengths are the same \(\ell_n\equiv \ell \) for all  \(n\). In that case
    \[ \xi(k)= -i 2^N e^{ikN \ell} \cos(k\ell)^{N-1} \sin(k \ell) .\]
    So the spectrum consists of two parts. First, the non-degenerate wave numbers 
    \(k= \frac{n \pi}{\ell}\) for non-negative integers \(n=0,1,\dots\). And second,
    the \(N-1\)-fold degenerate wave numbers \(k= \frac{\pi(2n+1)}{2\ell}\)
    for positive integers \(n=1,2,\dots\). In the latter case one can choose any
    subset of at least two bonds and construct eigenstates that are supported on the
    chosen bonds. For instance one may choose 
    \(\psi_1(x_1)= A \sin\left(\frac{\pi(2n+1)(\ell-x_1)}{2\ell}\right)\), 
     \(\psi_2(x_2)= -A \sin\left(\frac{\pi(2n+1)(\ell-x_2)}{2\ell}\right)\)
     and \(\psi_n(x_n) =0\) for all \(n\ge 2\) for an eigenstate that is supported on the
     bonds \(\me_1\) and \(\me_2\).
\end{example}

\subsubsection{The Barra--Gaspard approach: secular manifold}
\label{sec:BG}

\begin{figure}
  \centering
  \includegraphics[scale=1.0]{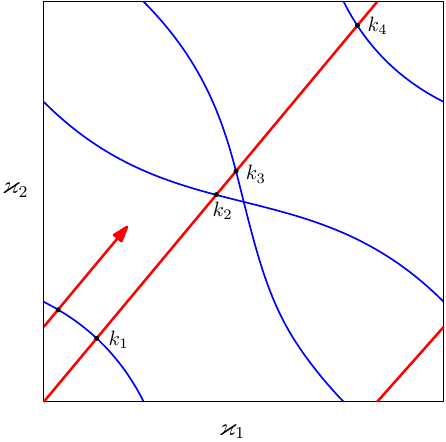}
  \caption{Flow (red) on the torus whose intersections with the
    secular manifold $\Sigma$ (in blue) generates the eigenvalues $k =
    \sqrt{E}$ of the underlying graph.  Note how the small
    nearest-neighbor spacing between $k_2$ and $k_3$ appears because
    the flow passes close to the singularity of $\Sigma$.}
  \label{fig:BGtorus}
\end{figure}

Originally due to Barra and Gaspard \cite{BarGas_jsp00}, the idea of
the \emph{secular manifold} is to observe that for scale-invariant
vertex conditions (such as Neumann-Kirchhoff or Dirichlet), 
the secular equation~\eqref{secular} has
the form
\begin{equation}
  \label{eq:secular_torus}
  0 = \xi(k) = F(k\ell_1, k\ell_2, \ldots, k\ell_B),
  \qquad B = |\Bset|.
\end{equation}
Moreover, the function
$F = F(\varkappa_1, \varkappa_2, \ldots, \varkappa_B)$ is
$2\pi$-periodic, so it can be thought of as a function on the
$|\Bset|$-dimensional torus
$\mathbb{T} = \left(\mathbb{R} / 2\pi\mathbb{R}\right)^{|\Bset|}$.  It
is an analytic function; furthermore, in the variables
$z_j = e^{i k \ell_j} = e^{i \varkappa_j}$ it is a polynomial
(\term{secular polynomial}).  Its zero set on the torus $\mathbb{T}$,
\begin{equation}
  \label{eq:secular_man_def}
  \Sigma := \left\{(\varkappa_1,\ldots,\varkappa_B) \in \mathbb{T} :
  F(\varkappa_1,\ldots,\varkappa_B)  = 0\right\},
\end{equation}
is called the \term{secular manifold} (even though it is not a
``manifold'' in the mathematical sense due to presence of
singularities).

%\begin{tcolorbox}
\begin{example}
  For the tadpole graph, see Example~\ref{ex:tadpole}, the function
  $F$ is
  \begin{equation}
    \label{eq:Ftadpole}
    F(\varkappa_1, \varkappa_2) = \frac13 \left(1-e^{i\kappa_2}\right)
    \left(3-	e^{i\varkappa_2}+e^{i2\varkappa_1}-3e^{i(2\varkappa_1+\varkappa_2)} \right);
  \end{equation}
  in terms of the variables $z_j$, the secular polynomial is
  \begin{equation}
    \label{eq:SecTadpole}
    \frac13(1-z_2)(3-z_2+z_1^2 - 3z_1^2z_2) = 0.
  \end{equation}
\end{example}
%\end{tcolorbox}

The eigenvalues $k = \sqrt{E}$ of the quantum graph can now be thought
of as the times of the intersection of $\Sigma$ by the flow
\begin{equation}
  \label{eq:BGflow}
  k \mapsto (k\ell_1, k\ell_2, \ldots, k\ell_B) \mod \pi \in \mathbb T.
\end{equation}
For rationally independent edge lengths, this flow is ergodic on torus
and the intersection points will lie densely on the secular manifold
$\Sigma$, distributed according to the cross-sectional measure,
defined according to the direction $(\ell_1,\ldots, \ell_B)$ and called
the \term{BG-measure} \cite{BerWin_tams10,CdV_ahp15,Alo_jam24}.

Barra and Gaspard observed that the small eigenvalue spacings are
generated by the singularities of the secular manifold $\Sigma$, see
Figure~\ref{fig:BGtorus}.  More generally, the nearest-neighbor
spacing distribution can be obtained by integrating along $\Sigma$
with the BG-measure \cite{BarGas_jsp00}.  Other statistics that can be
studied this way include the eigenvector statistics
\cite{BerWin_tams10,CdV_ahp15}, nodal statistics
\cite{AloBanBer_cmp17} and even resonance distribution
\cite{CdVTru_ahp18}.  In the study of Fourier quasicrystals, it has
been observed that the secular polynomials arising from quantum graphs
belong to a larger family of Lee-Yang polynomials
\cite{Rue_am10,KurSar_jmp20,AloCohVin_jfa24} which originate in the
study of partition functions in statistical mechanics
\cite{LeeYan_pr52}.

\subsubsection{Scattering states on an open quantum graphs}
\label{sec:scatter_states}

Next let us consider the positive energy states for open scattering  
quantum graphs \(\Gamma\)
which consists of \( \left|\Lset\right|\ge 1\) leads
and \(\left|\Bset\right| \ge 0 \) bonds.
In this case the number of incoming or outgoing plane wave amplitudes is 
\(N=\left| \Lset\right| + 2 \left| \Bset \right| \). Combining the 
incoming and outgoing amplitudes on the leads in a column vector
\( \mathbf{a}_{\Lset}^{\mathrm{in/out}}\) (analogous  to the bond in \eqref{bond_transport}).
In this context we define quantum map in terms of four blocks
\begin{equation}
\mathbf{U}(k)=
\begin{pmatrix}
 \mathbf{U}_{\Lset \Lset} & \mathbf{U}_{\Lset \Bset}\\
 \mathbf{U}(k)_{\Bset \Lset} & \mathbf{U}(k)_{\Bset \Bset}
\end{pmatrix}
=
\begin{pmatrix}
 \mathbf{S}_{\Lset \Lset} & \mathbf{S}_{\Lset \Bset}\\
\mathbf{T} \mathbf{S}_{\Bset \Lset} & \mathbf{T} \mathbf{S}_{\Bset \Bset}
\end{pmatrix} \ .
\end{equation}
One may then write \eqref{bond_transport} and \eqref{graph_scattering}
as
\begin{align}
	\begin{pmatrix}
	\mathbf{a}^{\mathrm{out}}_{\Lset}
	\\
	\mathbf{a}^{\mathrm{in}}_{\Bset}
	\end{pmatrix}
	=&
	\mathbf{U}(k)
	\begin{pmatrix}
	\mathbf{a}^{\mathrm{in}}_{\Lset}\\
	\mathbf{a}^{\mathrm{in}}_{\Bset}
	\end{pmatrix} \ .
\end{align}
This allows us to express the outgoing amplitudes on the leads
 in terms of the incoming amplitudes
 as
 \begin{equation}	
 	\mathbf{a}_\Lset^{\mathrm{out}}
 	= \mathbf{S}_{\Graph}(k) 
 	\mathbf{a}_\Lset^{\mathrm{in}}
 \end{equation}
 with a quantum graph scattering matrix
 %\GB{Are you sure the fraction notation is closer to our audience than
 %  the equivalent $(I-U)^{-1}$?}
\begin{equation}
\mathbf{S}_\Graph(k)=
\mathbf{U}_{\Lset \Lset}+
\mathbf{U}_{\Lset\Bset}
\frac{\mathbb{I}_{2\left|\Bset\right|}}{\mathbb{I}_{2\left|\Bset\right|}
-  \mathbf{U}(k)_{\Bset \Bset}}
\mathbf{U}(k)_{\Bset\Lset}\ .
\label{eq_scattering_matrix}
\end{equation}
%\GB{I made some modifications to the following paragraph, please
 % check} 
 One can show that \(\mathbf{S}_{\Graph}(k)\) is unitary using
that the quantum map \(\mathbf{U}(k)\) is unitary.  We have assumed
here that the matrix
\( \mathbb{I}_{2\left|\Bset\right|} - \mathbf{U}(k)_{\Bset \Bset}\) is
invertible and, while this is generically true (e.g. when the bond
lengths are chosen at random), it is straightforward to construct
graphs where this is violated, but only at isolated values of the wave
number \(k\).  One can show \cite[Lem.~2.5]{BanBerSmi_ahp12} that the
apparent poles in the above expression are in fact removable
singularities and \(\mathbf{S}_{\Graph}(k)\) is analytic on the real
line.  This is consistent with the physicist perspective where
unitarity of \(\mathbf{S}_\Graph(k)\) (which rules out divergence at
poles) just means current conservation.  In fact,
\( \mathbb{I}_{2\left|\Bset\right|} - \mathbf{U}(k)_{\Bset \Bset}\)
having a null space is know in mathematics literature as ``absence of
unique continuation'' or existence of ``inner solutions'' (see, e.g.,
\cite[Thm.~3.8]{BooFur_tjm98}).  On the physical side, it leads to
appearance of perfect scars (see Example~\ref{ex:scatt_loop}) and
topological resonances which will be discussed in
\Cref{sec:topological_resonances} below.

% The graph scattering
% matrix \(\mathbf{S}_\Gamma(k)\) allows us
% to think of the internal part of the graph as a black box and focus on the behaviour on the leads.\\
If one seeks information for the internal structure of the scattering
state one can use the relation
\begin{equation}
	\mathbf{a}_{\Bset}^{\mathrm{in}}= \mathbf{R}_{\Gamma}(k) 
	\mathbf{a}^{\mathrm{in}}_\Lset
\end{equation}
where
\begin{equation}
\mathbf{R}_\Graph(k)=
\frac{\mathbb{I}_{2\left|\Bset\right|}}{\mathbb{I}_{2\left|\Bset\right|}
-  \mathbf{U}(k)_{\Bset \Bset}}
 \mathbf{U}(k)_{\Bset\Lset}
\end{equation}
which expresses internal plane wave amplitudes in terms of
incoming amplitudes.

%\begin{tcolorbox}
\begin{example}
\label{ex:scatt_loop}
 	Let us consider scattering from a loop (this is again a tadpole graph 
 	as in Example~\ref{ex:tadpole} with the difference that the
        dangling bond edge is replaced 
 	by a lead edge). 
 	This is described by a 
 	simple open graph with one lead \(\me_1\)  and one bond
 	\(\me_2\) of length \(\ell\)
 	that forms a loop that starts and ends at the same vertex where the lead is attached
 	\(o(\me_1)=o(\me_2)=t(\me_2)\). Note that this graph is an open variant of the tadpole 
 	graph discussed above.
 	In this case we have
 	\begin{align}
 	\mathbf{S}_{LL}=&
 	-\frac{1}{3},\quad &
 	\mathbf{S}_{LB}=&
 	\begin{pmatrix}
 	\frac{2}{3} & \frac{2}{3}
 	\end{pmatrix},\quad &
 	\mathbf{S}_{BL}=& \mathbf{S}_{LB}^T=
 	\begin{pmatrix}
 	\frac{2}{3}\\
 	\frac{2}{3}
 	\end{pmatrix},\quad \text{and} &
 	\mathbf{S}_{BB}=&
 	\begin{pmatrix}
 	\frac{2}{3} & -\frac{1}{3}\\
 	-\frac{1}{3} & \frac{2}{3}
 	\end{pmatrix}.
 	\end{align}
 	It is straightforward to deduce the graph scattering matrix
 	\begin{equation}
 	\mathbf{S}_\Graph(k) =e^{i k \ell}
 	\frac{3  - e^{-ik\ell}}{3-e^{ik\ell}}, \quad \text{and}
 	\quad \mathbf{R}_\Graph(k)=
 	\frac{2 e^{i k \ell}}{ 3-e^{ik\ell} }
 	\begin{pmatrix}
 	1\\
 	1
 	\end{pmatrix} .
 	\end{equation}
 	Note that 
 	\begin{equation}
 	\left( \mathbb{I}_2 -  e^{i k \ell}\mathbf{S}_{BB}\right)^{-1}
 	=
 	\frac{3}{\left(1-e^{i k \ell} \right) \left(3-e^{ik\ell} \right)}
 	\begin{pmatrix}
 	1-\frac{2e^{ik\ell}}{3}&
 	- \frac{e^{ik\ell}}{3}\\
 	- \frac{e^{ik\ell}}{3} & 
 	1-\frac{2e^{ik\ell}}{3}
 	\end{pmatrix}
 	\end{equation}
 	has poles at  whenever \(k\) is an integer multiple of \(\frac{2\pi}{\ell}\)
 	(in that case \(e^{i k \ell}=1\). These poles disappear in \(\mathbf{S}_\Graph(k) \)
 	and \(\mathbf{R}_\Graph(k)\). 
 	We can relate this to the perfect scars we noticed above when discussing the closed
 	tadpole graph (see Example~\ref{ex:tadpole}). 
 	The same perfect scars reappear here as \textit{bound states
 	in the continuum}. For \(k \neq \frac{2\pi n}{\ell} \)for any positive integer one 
 	only has the continuum of scattering states. When 
 	 \(k = \frac{2\pi n}{\ell} \) one has the scattering state
 	 \begin{equation}	
 	 	\psi^{\mathrm{cont}}_1(x_1)=
 	 	A \cos\left(\frac{2 \pi n}{\ell} x_1\right)
 	 	\quad \text{and} \quad
 	 	\psi^{\mathrm{cont}}_2(x_2)=
 	 	A \cos\left(\frac{2 \pi n}{\ell} x_2\right)
 	 \end{equation}
 	 %with incoming amplitude \( a^{\mathrm{in}_{e_1}= \frac{A}{2}} \)
 	 \textit{and} an orthogonal bound (normalizable) state
 	 \begin{equation}	
 	 	\psi^{\mathrm{bound}}_1(x_1)=0
 	 	\quad \text{and} \quad
 	 	\psi^{\mathrm{bound}}_2(x_2)=
 	 	\sqrt{\frac{2}{\ell}} \sin\left(\frac{2 \pi n}{\ell} x_2\right) .
 	 \end{equation} 
\end{example}
%\end{tcolorbox}

%\begin{tcolorbox}
\begin{example}
   Next let us consider scattering in an open star graph 
   with \(N_L\equiv \left|\Lset\right|\ge 1\)
   leads \(l_{1},\dots l_{N_L}\)
   and \(\left|\Bset\right|\ge 0\) dangling bonds
  \(b_{1},\dots b_{|\Bset|}\) of lengths
   \(\ell_{1},\dots \ell_{M_B}\).
   We write \(N=N_L+N_B=\left|\Eset\right| \) for the total number
   of edges
   There is one central vertex which we take to be the origin of all edges
   \(o(l_1)=o(l_2)=\dots=o(b_1)=\dots=o(b_{|\Bset|})\).
   For \(|\Bset|=0\)  the graph scattering matrix is
   just the vertex scattering matrix and is thus a trivial case.
   In this case one finds
   \begin{align}
   	\mathbf{S}_{LL}=& -\mathbb{I}_{N_L}+\frac{2}{N} \mathbb{E}_{N_L},
   	\quad&
   	\mathbf{S}_{LB}=& 
   	\begin{pmatrix}
   	0_{N_L \times N_B} &
   	\frac{2}{N} \mathbb{E}_{N_L \times N_B}
   	\end{pmatrix},
   	\quad &
   	\mathbf{S}_{BL}=& 
   	\begin{pmatrix}
   	\frac{2}{N} \mathbb{E}_{N_B \times N_L}\\
   	0_{N_B\times N_L}
   	\end{pmatrix}, 
   	\quad \text{and} &
   	\mathbf{S}_{BB}=&
   	\begin{pmatrix}
   	 0_{N_B} & -\mathbb{I}_{N_B}+\frac{2}{N} \mathbb{E}_{N_B}\\
   	\mathbb{I}_{N_B} & 0_{N_B}
   	\end{pmatrix}
   \end{align}
   and the graph scattering matrix is
   \begin{equation}
   \mathbf{S}_\Gamma(k)
   = -\mathbb{I}_{N_L} + \frac{2}{N_L -i \sum_{n=1}^{N_B} 
   \tan(k\ell_{b_{n}})}\mathbb{E}_{N_L},
   \quad\text{and}\quad
   \mathbf{R}(k)=
    \frac{2}{N_L -i \sum_{n=1}^{N_B} 
   \tan(k\ell_{b_{n}})} 
   \begin{pmatrix}
   \frac{\mathbf{T}_{\mathrm{red}}(k)}
	{\mathbb{I}_{N_B} + \mathbf{T}_{\mathrm{red}}(k)^2}
   \mathbb{E}_{N_B\times N_L}\\
    \frac{\mathbf{T}_{\mathrm{red}}(k)^2}
	{\mathbb{I}_{N_B} + \mathbf{T}_{\mathrm{red}}(k)^2}
   \mathbb{E}_{N_B\times N_L}
   \end{pmatrix}
   \end{equation}
\end{example}
%\end{tcolorbox}

%\GB{Do you want to explain how the eigenstates can be detected as
 % solutions of $\det(\mathbf{S}_\Gamma(k) - \mathbb{I}_{N_L})$ --- for
 % NK conditions at the lead attachment points?  This can then be
 % related to DtN map formula \eqref{eq:secular_dtn}.}
  %\SG{Maybe that's a good idea - inside -outside duality was not as heavily used in quantum 
  %graphs as for billiards but it fits well with DtN and maybe that's where it should go? I would 
  %defer this till we have a complete draft}
%  \SG{After reading this again I believe it should really be added -- but it fits better in the DtN 
%chapter.}

\subsubsection{More general graph structures with infinite number of edges}

We have described the scattering approach for graphs with a finite
number of edges in detail above. It is straight forward to come up
with relevant graph structures that have an infinite number of
edges. To give one example, one may take a compact graph and copy it
in a periodic fashion, connecting the copies to create a periodic
lattice structure (see \cite[Chap. 4]{BerKuc_graphs} for a complete
setup on graphs and \cite{Kuc_bams16} for more general constructions).
One obtains a band-gap spectrum, which can be engineered to have gaps
in specific locations
\cite{AvrExnLas_prl94,Exn_prl95,DoKucOng_emsscr17,ExnTur_jpa17}, have
Dirac points \cite{KucPos_cmp07,Do_amp15} or studied from the point of
view of spectral statistics \cite{BanBer_prl13}.

Then one may add disorder to a periodic structure that breaks the
periodicity of the lattice, for instance to study Anderson
localization on quantum graphs \cite{SchSmi_prl00,DamFilSuk_ma20}.  In
a similar way one can construct tree-like structures or graphs
modeled on fractals
\cite{AloKelTep_jpa16,ExnKosMalNei_ahp18,KostenkoNicolussi_book}.  We will not dive
deeper into the construction of the scattering approach for these
structure here but we will come back to one related topic -- quantum graph models 
for metamaterials  \cite{lawrie2022quantum,lawrie2024application,lawrie2025nondiffracting} in  
\Cref{sec:metamaterials}.

\subsection{Quantum-classical correspondence: semiclassical sums over trajectories and 
periodic orbits}

Quantum graphs have become a paradigm for quantum chaos after Kottos and Smilansky showed
in a series of papers  
\cite{KotSmi_prl97,KotSmi_ap99,KotSmi_prl00,KotSmi_jpa03}
that semiclassical approaches 
to quantum chaos
(see \cite{tomsovic_2026, sieber_muller_2026, novaes_2026} in this volume for overviews of 
various aspects of the 
semiclassical approach to Hamiltonian systems)
have a close analogy on quantum graphs -- with two two major differences:
the expressions are formally exact and the underlying dynamics is far less complex than a chaotic
Hamiltonian flow. Indeed the dynamics is just free motion along each edge until a vertex is hit
when one is either reflected or scatters into one of the other adjacent edges. 
For this purpose the quantum dynamics is best described in terms
of the quantum map \( \mathbf{U}(k)\) for closed graphs. 
In this way the dynamics is a dynamics in the space of 
\(2\left|\Bset\right|+\left|\Lset\right|\) quantum amplitudes on directed bonds and leads.
A (finite) \textit{trajectory} or \textit{walk} \(\theta\) 
of \textit{topological length} \(n\) is a sequence of \(n+1\) 
directed edges  
\(\theta= \left(\me_{n \, \sigma_{n}},\dots,\me_{1\, \sigma_1},\me_{0,\ \sigma_0} \right) \) 
that are all following each other  
 \(\me_{m\, \sigma_m} \rightarrow \me_{m+1\, \sigma_{m+1}}\).
%\(t(e_{m\, \sigma_m})=o(e_{m+1\, \sigma_{m+1}})\).
A \textit{closed trajectory} starts and ends o the same directed edge
\( \me_{n\, \sigma_n}=\me_{0\, \sigma_0}\).
The \textit{quantum amplitude} of the trajectory
\begin{equation}
	\prod_{m=1}^{n} \mathbf{U}(k)_{\me_{m\, \sigma_m} \me_{m-1\, \sigma_{m-1}}}
	= A_{\theta } e^{i k L_{\theta}} \quad
	\text{with}\quad 
	A_{\theta} =
	\prod_{m=1}^{n} \mathbf{S}_{\me_{m\, \sigma_m} \me_{m-1\, \sigma_{m-1}}}
	\quad \text{and} \quad L_\theta= \sum_{1\le m\le n: \me_m \in \Bset} \ell_{\me_{m}}
\end{equation}
 is the product of the corresponding
transition amplitudes in the quantum map. 
The metric length \( L_\theta\) of the trajectory excludes the initial edge \(e_0\)
(in this context one considers \(t(e_0)\) as the starting point of the trajectory) and the prime
indicates that any edges of infinite length (leads) are excluded. For elevant trajectories
on open graphs
leads only enter either as final edge \(e_n\) or starting edge  \(e_0\) (or both) -- while all
edges in between are always bonds.

\subsubsection{The classical map}
\label{sec:classicalmap}

Quantum-classical correspondence on quantum graphs 
is obtained by defining
a \textit{classical map} \(\mathbf{M}\) that corresponds to the quantum 
map \(\mathbf{U}(k)\).
This is a Markov chain that evolves probabilities on the space of directed edges.
As a map \(\mathbf{M}\)
is defined by a  \(N\equiv 2\left|\Bset \right|+ \left|\Lset\right| \)-dimensional square matrix
of transition amplitudes
that is obtained by taking the absolute squares 
\begin{equation}
\mathbf{M}_{e'_{\sigma'} e_{\sigma}}=
\left|\mathbf{U}(k)_{e'_{\sigma'} e_{\sigma}}\right|^2
= \left|\mathbf{S}_{e'_{\sigma'} e_{\sigma}}\right|^2
\end{equation}
of the quantum transition
amplitudes in the quantum map. Note that the quantum map contains 
energy (wave number)  dependent phase factors while the classical map does not depend
on the energy at all.
If we denote the probabilities at a time step \(n\) by\(p_{e_\sigma}(n)\) and combine them
into a column vector \(\mathbf{p}(n)\) then the dynamics may be written as
\( \mathbf{p}(n+1)= \mathbf{M} \mathbf{p}(n)\).  
This is just a simple finite dimensional Markov chain defined by a 
stochastic matrix.
Indeed the matrix \(\mathbf{M}\) is \textit{bi-stochastic} which means
that sums over rows or columns all add up to one. That is for 
any fixed directed edge \(e'_{\sigma'}\) one has
\begin{equation}
	\sum_{e_\sigma} \mathbf{M}_{e'_{\sigma'} e_\sigma}=
	\sum_{e_\sigma} \mathbf{M}_{e_{\sigma} e'_{\sigma'}}=1\ .
\end{equation}
This property follows directly from the fact that the quantum map \(\mathbf{U}(k)\) is unitary.
It implies that \(\mathbf{p}^{\mathrm{inv}}= \frac{1}{N} \mathbb{E}_{N\times 1}\)
is an invariant probability distribution
\begin{equation}
  \mathbf{M} \mathbf{p}^{\mathrm{inv}}= \mathbf{p}^{\mathrm{inv}} \ ,
\end{equation}
and, in particular, $\mathbf{M}$ has an eigenvalue 1.  Furthermore,
all other eigenvectors of $\mathbf{M}$ are orthogonal to
$\mathbf{p}^{\mathrm{inv}}$ (which is not automatic since $\mathbf{M}$
is not symmetric).  Any eigenvalue
\(\lambda\) of \(\mathbf{M}\) has the property
\(\left| \lambda\right|\le 1\).

There are two candidates for the notion of ``chaotic'' classical map;
both are relevant to spectral statistics, see the discussion in
Section~\ref{sec:tanner}.  The Markov chain $\mathbf{M}$ is
\emph{ergodic} if the eigenvalue $\lambda=1$ has
multiplicity\footnote{In the setting of Markov chains, the geometric
  and algebraic multiplicites of the eigenvalue 1 are equal by
  \cite[Thm~3.2.5.2]{HornJohnson}.} one.  Equivalently, we have a
non-zero \term{spectral gap}
\(\Delta \lambda = \min \left| 1-\lambda\right|\),
where the minimum is taken over all eigenvalues of \(\mathbf{M}\) on
the subspace orthogonal to $\mathbf{p}^{\mathrm{inv}}$.  If the Markov
chain is ergodic,
\begin{equation}
  \label{eq:ergodic_thm}
  \lim_{n \to \infty}\frac{1}{n} \sum_{m=1}^n \mathbf{M}^{m} \mathbf{p}(0)
  = \mathbf{p}^{\mathrm{inv}},
\end{equation}
for any initial distribution $\mathbf{p}(0)$.

A stronger property of the Markov chain $\mathbf{M}$ is being \term{mixing}: all eigenvalue
other than the unique eigenvalue 1 are \(\left| \lambda\right|< 1\).
Equivalently, the \emph{absolute spectral gap}
 \(\Delta^*\lambda = \min (1-|\lambda|)\) is non-zero (as before, the minimum is
taken over all eigenvalues other than 1).  For a mixing Markov chain,
we have a stronger version of \eqref{eq:ergodic_thm},
\begin{equation}
  \label{eq:convergence_thm}
  \lim_{n \to \infty} \mathbf{M}^{n} \mathbf{p}(0)
  = \mathbf{p}^{\mathrm{inv}},
\end{equation}
for any initial distribution $\mathbf{p}(0)$.  Curiously, presence or
absence of the time average in
\eqref{eq:ergodic_thm}-\eqref{eq:convergence_thm} is echoed in the
spectral statistics of $\mathbf{U}(k)$, see
\eqref{eq:K_time_averaged} and the discussion around it.

%\GB{If you like the version above, please remove this paragraph}
%The classical map is chaotic if the
%corresponding Markov chain is ergodic.  In the present context that means
%\begin{equation}
%\lim_{n \to \infty}\frac{1}{n} \sum_{m=1}^n \mathbf{M}^{m} \mathbf{p}(0)
%= \mathbf{p}^{\mathrm{inv}} .
%\end{equation}
%Alternatively one may characterise ergodic maps by two properties:
%first, the invariant probability distribution is unique (the
%eigenvalue \(1\) is non-degenerate), and second, the \textit{spectral
%  gap} \(\Delta \lambda = \mathrm{min}_{\lambda \neq 1} \left|
%  1-\lambda\right| >0\)
%(where the minimum is over all eigenvalues
%apart the unique eigenvalue \(1\)).

%\GB{I lightly edited this paragraph to fit the dichotomy.}
It can be shown that ergodicity holds for any finite connected quantum
graph (even beyond the Kirchhoff-Neumann matching conditions if one
replaces the topological notion of connectivity by a dynamical one).
Mixing is a little more subtle: for example, a Neumann interval (or,
more generally, a tree) is not mixing because the edge dynamics is
bipartite.  The iteration time scale when ergodicity (corresp.~mixing)
sets in is given by \(n > \frac{1}{\Delta \lambda}\)
(corresp.~\(n > \frac{1}{\Delta^* \lambda}\)), see,
e.g. \cite[Chap.~12]{LevinPeres_MarkovChains}.

One can view the quantum map \(\mathbf{U}(k)\) as a quantization of
the chaotic classical map \(\mathbf{M}\). While the classical map
cannot be gained from the quantum map in a semiclassical limit as in
other paradigmatic systems of quantum chaos such as quantum billiards,
kicked tops, quantum cat maps or the kicked rotator we will see that
standard semiclassical approximations for the latter can be given a
formally exact meaning in quantum graphs -- with an analogous relation
between quantum amplitudes \(A_{\theta}e^{ik L_\theta}\) and
corresponding classical amplitudes \(\left| A_\theta\right|^2\)
for a trajectory \(\theta\)  in the system. \\
In the semiclassical approach to quantum chaos one considers quantum
signatures of chaos in the asymptotic (formal) limit \(\hbar \to 0\).
The relevant quantum mechanical correlation functions are based on a
number of quantities such as propagators, scattering matrices, Greens
functions, density of states, and related correlations functions.
Using the WKB approximations these quantities can be written as sums
over trajectories \(\theta\) of the corresponding classical dynamics
with amplitudes
\(\propto A_\theta e^{i \int_\theta \mathbf{p} d\mathbf{x}/\hbar} \)
where the amplitude \(A_\theta\) squares to a classical decay rate of
the underlying flow (as described by a classical Lyapunov
exponent). This is analogous to the quantum graph amplitudes where the
classical decay is described by the classical map.  With the relation
\(k=p/\hbar\) between quantum wave number and momentum one finds that
the phase \(k L_\theta =p L_\theta/\hbar \) in quantum graph
amplitudes is completely analogous to the general semiclassical case.
Below we will state formally exact expressions for the scattering
matrix and the density of states as sums over quantum amplitudes of
trajectories.  In both cases the expressions go back to the seminal
work of Kottos and Smilansky that introduced quantum graphs as a
paradigm for quantum chaos
\cite{KotSmi_prl97,KotSmi_ap99,KotSmi_prl00,KotSmi_jpa03}.

\subsubsection{Scattering as sum over trajectories}

Let us consider an open graph with the 
scattering matrix \(\mathbf{S}_\Gamma(k)\) defined in \eqref{eq_scattering_matrix}.
It describes how incoming plane waves on the leads 
are scattered through the graph system and result
in a particular configuration of outgoing plane waves on the leads.
Fixing two leads \(\me^{\mathrm{in}}\) and \(\me^{\mathrm{out}}\) one may write
the corresponding matrix element of the scattering matrix as
\begin{equation}
\mathbf{S}_\Gamma(k)_{\me^{\mathrm{out}} \me^{\mathrm{in}}}=
\sum_\theta A_\theta e^{i k L_\theta}
\label{eq_semiclassical_scattering}
\end{equation}
where the sum is over all (finite) trajectories 
\(\theta= \left(\me_{n\, \sigma_n},\dots, \me_{0\, \sigma_0}  \right) \)
of (topological) length \(n\ge 1\) with the following restrictions.
First, the trajectory \(\theta\) has to start and end at the corresponding leads
\(\me_{0\, \sigma_0}= \me^{\mathrm{in}}_-\) 
and \(\me_{n\, \sigma_n}= \me^{\mathrm{out}}_+\). Second, for \(1\le m \le n-1\)
the directed
edges \(\me_{m\, \sigma_m}\) along \(\theta\) are directed bonds.\\
The derivation of \eqref{eq_semiclassical_scattering}
is straight forward. 
Starting from the scattering matrix \eqref{eq_scattering_matrix}
it amounts to writing 
\begin{equation}
\frac{\mathbb{I}_{2|\Bset|}}{\mathbb{I}_{2 |\Bset|} - \mathbf{U}(k)_{\Bset\Bset}}
= \sum_{m=0}^\infty \left(\mathbf{U}(k)_{\Bset\Bset}\right)^m
\end{equation}
as a geometric series.
The matrix \(\mathbf{U}(k)_{\Bset\Bset}\)
is a quadratic block in the unitary matrix \(\mathbf{U}(k)\). This implies that all
eigenvalues of \(\mathbf{U}(k)_{\Bset\Bset}\) have modulus smaller or equal to one.
Whenever all eigenvalues of have modulus smaller than one then the
expansion into a geometric series and with it the expression
 \eqref{eq_semiclassical_scattering} converge absolutely.
 \begin{remark}
 We have already seen examples (tadpole/lasso graph and star graphs) where bound states
 may exist that correspond to eigenvalues equal to one.
 So in general one should not expect the sum over trajectories to converge absolutely 
 even if it may do so for many specific energy intervals (wave numbers).
The scattering matrix as expressed in
 \eqref{eq_scattering_matrix} can be given an exact meaning
as one can show that poles in
\(\frac{\mathbb{I}_{2|\Bset|}}{\mathbb{I}_{2 |\Bset|} - \mathbf{U}(k)_{\Bset\Bset}}\)
are cancelled in the full expression for the scattering matrix -- in the sum
over trajectories this implies that there are many cancellations due to interferences of
terms. This requires either a specific ordering of terms or a regularization
(e.g. by replacing \(k \to k+i\epsilon\) and taking the limit \(\epsilon \to 0^+\)).
\end{remark}

\subsubsection{The Roth-Kottos-Smilansky trace formula and periodic orbit
  theory}
\label{sec:trace_formula}

%\GB{Sorry, but we should not be calling it Kottos--Smilansky as
%  precisely this formula was established by Roth and we know it...
%  Roth--Kottos--Smilansky?}

For chaotic Hamiltonian systems the spectra of the corresponding
quantized system can be described by the semiclassical Gutzwiller
trace formula.  It expresses the density of states
\(\rho(E) =\sum_n \delta(E-E_n)\) (where \(E_n\) are the energy
eigenvalues) as a sum over periodic orbits (equivalence classes of
closed phase space trajectories) \cite{Gutzwiller_book,QSoC}. 
% \SG{refer to other chapter by Sebastian and Martin} 
The Gutzwiller trace formula is based on the
WKB approximation and as such on asymptotic approximations that
formally take \(\hbar \to 0\).

Remarkably, for quantum graphs a similar trace formula is exact: it
does not require any asymptotic limit.  This formula has been derived
by Roth \cite{Rot_crasp83,Rot_incol84} and, independently, by Kottos
and Smilansky \cite{KotSmi_prl97,KotSmi_ap99}.  Before stating the
result let us define a \textit{periodic orbit}
\( p=\overline{\me_{n_p\, \sigma_{n_p}}\dots \me_{2\, \sigma_2} \me_{1\,
    \sigma_1}} \equiv \overline{ \me_{{n_p}-1\, \sigma_{{n_p}-1}} \dots
  \me_{1 \sigma_1} \me_{{n_p}\, \sigma_{n_p}} } \) of (topological) length
\(n_p\ge 1\) as an equivalence class of closed trajectories that go
through the same set of directed edges in the same order with the only
difference being what directed edge is taken to be the start. For each
periodic orbit we define the quantum amplitude
\(A_p e^{i k L_p}= A_{\theta}e^{i k L_{\theta}}\) to be the same as
the one of the corresponding closed trajectories.  If a periodic orbit
\(p\) of length \(n_p\) is a repetition of a shorter periodic orbit
\(p'\) with length \(n_{p'}=n_p/r\) (that is the same sequence of
directed edges is revisited \(r\) times) one can write
\( A_p e^{ik L_p}=A_{p'}^r e^{i k r L_{p'}} \).  A \textit{primitive
  periodic orbit} is defined as a periodic orbit that is not a
repetition of a shorter periodic orbit. If \(p\) is a periodic orbit
that is a repetition of the primitive periodic orbit \(p'\) then we
define the repetition number \(r_p\) of
\(p\) as the ratio \(r_p= n_p/ n_{p'}\).\\
The first important observation is that the secular function can
be written as infinite product over primitive periodic orbits
\begin{equation}
\xi(k)=
\det
\left(\mathbb{I}_{2\left|\Bset\right|}- \mathbf{T}(k) \mathbf{S}  
\right)
= e^{\mathrm{tr}\, 
\log\left( \mathbb{I}_{2\left|\Bset\right|}   -   \mathbf{T}(k) \mathbf{S}    \right)}
= 
e^{-\lim_{\epsilon \to 0^+}\sum_{n=1}^\infty \frac{e^{-n \epsilon}}{n} \mathrm{tr}\, 
\mathbf{U}(k)^n }
= \lim _{\epsilon\to 0^+}
\prod_{p} \left(1- A_p e^{-n_p \epsilon + ik L_p}\right)
\end{equation}
where all primitive periodic orbits of lengths \(n_p \ge 1\) contribute.
In the following we will omit the limit \(\epsilon \to 0 \).
Using Cauchy's argument principle this leads directly to the %Kottos-Smilansky 
Roth-Kottos-Smilansky trace
formula
\begin{equation}
\rho(k)= \sum_n
\delta(k-k_n)
= \frac{L_\Gamma }{\pi}
- \frac{1}{\pi} \frac{d}{dk} \mathrm{Im}\, \log\, \xi(k)
 =  \frac{L_\Gamma }{\pi}
 + \mathrm{Re} \sum_p   \frac{L_p}{\pi}
 \sum_{r=1}^\infty  A_p^r e^{ik r L_p}
 \label{eq:Kottos-Smilansky}
\end{equation}
for the density of states of a quantum graph.
Here \(L_\Gamma= \sum_{e \in \Bset} \ell_e\) is the total metric length of the graph
and the sum is over all primitive periodic orbits \(p\) and their repetitions \(r\).
The trace formula expresses the density of states in terms of a mean value (known as Weyl
part) \(\frac{L_\Gamma}{\pi}\) which gives the inverse mean spacing between 
the wave numbers in the spectrum and fluctuations around the mean that are 
expressed as a sum over amplitudes of periodic orbits.
This closely corresponds structurally to  the Gutzwiller trace formula in Hamiltonian flows 
%\SG{(add reference to other chapter)} 
with the difference that it is an exact formula
that can be given rigorous meaning while the Gutzwiller trace formula is a 
semiclassical approximation.
The impact of the of the quantum graph
trace formula in Quantum Chaos 
after the work of Kottos and Smilansky is mainly based on this
analogy and the fact that information about long periodic orbits in graphs can be produced with
relative ease.

To give this formula a precise mathematical meaning, it is essential
to recognize that both sides of the equation are not functions but are
in fact \textit{tempered distributions}
\cite{Strichartz_distribution_theory}. The equality is therefore to be
understood in the distributional sense, meaning that both sides yield
the same result when applied to a suitable test function.

Let $h(k)$ be a test function from the Schwarz space
$\mathcal{S}(\mathbb{R})$. A more fundamental and symmetric form of
the trace formula arises when we consider the full spectrum of roots
of the secular equation $\det
\left(\mathbb{I}_{2\left|\Bset\right|}- \mathbf{T}(k) \mathbf{S}  
\right)=0$.  Let $\{k_n\}_{n=-\infty}^\infty$ be this set, where each
root is listed with its multiplicity.  We define the full spectral
measure as $\mu(k) = \sum_{n=-\infty}^\infty \delta(k-k_n)$.  The corresponding 
geometric side of the formula no longer requires taking the real part. 
The symmetrized trace formula is:
\begin{equation}
  \label{eq:full_trace_formula}
  \sum_{k_n \in \mathbb{R}} h(k_n) = \frac{L_\Gamma}{\pi} \hat{h}(0)
  + \sum_{p} \frac{L_p}{\pi} \sum_{r \neq 0} A_p^r \hat{h}(rL_p),
  \qquad\text{where }
  \hat{h}(\xi) = \int_{-\infty}^\infty e^{-i \xi x} h(x)dx,
\end{equation}
and where the sum over repetitions $r\in \mathbb{Z}\setminus \{0\}$
now runs over all non-zero integers. This equation is an analogue of
the Poisson summation formula
\cite[Sec.~7.3]{Strichartz_distribution_theory}, and it is through
this connection that the theory of quantum graphs have provided some
examples of Fourier quasicrystal with peculiar properties, see
Section~\ref{sec:FQ}.

%%%%%%%%%%%%%%%%%%%%%%%%%%%%%%%%%%%%%%% 
\subsection{Generalizing matching conditions}

We have defined quantum graphs by requiring the wave function to be
continuous at the vertices.  We then continued to develop most
of the theory for the special case of Neumann--Kirchhoff conditions
which can be considered as the canonical case for a quantum
graph. There are however many ways to extend the model.  Some
based on mathematical considerations, other on practical
considerations of what one may encounter in a lab.

\subsubsection{Self-adjoint matching conditions}
\label{sec:selfadjoint}
On the mathematical side one considers the most general type of
matching conditions which render the metric Laplacian
\eqref{eq:laplace_def} and related operators (including potentials and
magnetic fields) self-adjoint.  There are several equivalent
formulations \cite[Thm.~1.4.4]{BerKuc_graphs}, and we follow here the
original formulation as derived by Kostrykin and Schrader
\cite{KosSch_jpa99}.  It is important to stress that Cheon, Exner and
Turek have demonstrated that \emph{all} types of vertex conditions
considered in this section can be realized as limits of subgraphs with
short edges and the more physical $\delta$-type conditions
\cite{CheExn_jpa04,CheExnTur_anp10}.

Consider a vertex $\mv$, which we assume to
be the origin ($x=0$) of the edges $\me_{j_1},\ldots,\me_{j_m}$ and
the terminus ($x=\ell$) of the edges $\me_{j_1'},\ldots, \me_{j'_n}$.
We introduce two vectors
\begin{equation}
  \label{eq:DNdata}
  F_\mv =
  \begin{pmatrix}
    \psi_{j_1}(0) \\ \vdots\\ \psi_{j_m}(0) \\
    \psi_{j_1'}(\ell_{j_1'}) \\ \vdots \\ \psi_{j_n'}(\ell_{j_n'}) 
  \end{pmatrix},
  \qquad\text{and}\qquad
  F'_\mv =
  \begin{pmatrix}
    \frac{d}{dx}\psi_{j_1}(0) \\ \vdots \\ \frac{d}{dx}\psi_{j_m}(0) \\
    -\frac{d}{dx}\psi_{j_1'}(\ell_{j_1'}) \\ \vdots\\  -\frac{d}{dx}\psi_{j_n'}(\ell_{j_n'})
  \end{pmatrix},
\end{equation}
known as the \term{Dirichlet data} and \term{Neumann data} correspondingly.
The self-adjoint matching conditions have the following form at every
vertex $\mv$:
\begin{equation}
  \label{eq:sa_matching}
  A_\mv F_\mv + B_\mv F'_\mv = 0,
\end{equation}
where the $(m+n)\times(m+n)$ matrices $A_\mv$ and $B_\mv$ are such that
the concatenated matrix $(A_\mv\ B_\mv)$ has maximal rank and the matrix
$A_\mv B_\mv^\dagger$ is Hermitian.

For most matching conditions (including the $\delta$-type conditions
\eqref{eq:cond_current} with $\alpha_\mv \neq 0, \infty$),
the corresponding vertex scattering matrices
\(\boldsymbol{\sigma}^{v}(k)\) depend explicitly on the wave number
\(k\).  The edge scattering matrix $\mathbf{S}$ appearing in the
secular equation \eqref{quantum_map}-\eqref{secular} is no-longer constant.
The scattering approach can still be developed in
analogy to the derivation above.  Where derivatives of the quantum map
are taken as, for instance, in \eqref{eq:Kottos-Smilansky} additional
terms appear that stem from the \(k\)-dependence of the matching
conditions.  The rigorous derivation of the Roth--Kottos--Smilansky
trace formula for general self-adjoint matching conditions was done in
\cite{BolEnd_incol08,BolEnd_ahp09}.

%%%%%%%%%%%
\subsubsection{Vertices as wave-scatterers}

In physics and engineering a more phenomenological approach has often
been pursued: directly prescribing the vertex scattering matrices \(
\boldsymbol{\sigma}^{v}(k)\) --- which may depend on the wave number \(k\).
From an experimental perspective
this scattering approach is perfectly adapted for any network of
effectively one dimensional carriers of waves --- quantum, electrical,
optical, mechanical or acoustic ones.  In this approach the
vertices may themselves be composite physical systems with the vertex
scattering matrix describing their response to incoming waves.
While there is no self-adjoint operator that describes the vertices,
this is not unphysical; rather, it means that part of the Hilbert space is
`hidden' inside the vertex.

To give an example consider a scattering graph with scattering matrix
\(\mathbf{S}_\Graph(k) \) as we have developed above. We can now
replace all the internal structure of the scattering matrix by a
single effective vertex where matching conditions are given by
\(\mathbf{S}_\Graph(k) \).  
If we think of an hypothetical experiment where the internal structure of
\(\Graph\) is not known an can only be measured from the outside it does make sense
to think of the whole internal graph structure as one effective vertex with a vertex
scattering matrix \(\mathbf{S}_\Graph(k) \) that can be measured.
One should just be aware that the presence of such effective vertices in time-dependent
wave functions will apparently break probability conservation
on the visible edges as some probability can flow
to and from the hidden interior edges.
One should also not be surprised that eigenfunctions in this case may not
form an orthonormal basis (unless their full structure on the interior 
edges is considered as well).
Replacing subgraphs by effective vertices with the corresponding effective vertex scattering 
matrix can be a useful approach practically, for instance in computer simulations.
%\GB{References? I know of ``resonant gap opening'', but they
%  insert a DtN map rather than S (which is of course equivalent)}.
  %\SG{I have reformulated the above to make the point - 
  %references are not really needed here.}

In quantum chaos-related theoretical applications, this freedom has often been used
to choose a \emph{constant vertex scattering matrix}
\(\boldsymbol{\sigma}^{v}\) that has preferable physical
characteristics -- such as scattering to all connected edges with same
or similar probabilities \cite{HarSmiWin_jpa07} (while the
Neumann-Kirchhoff conditions have a a strong preference for scattering
back if the degree is high).

\subsubsection{Connection between the two approaches}

The apparent differences between the two approaches, the self-adjoint
vertex conditions \eqref{eq:sa_matching} with constant $A_\mv$ and
$B_\mv$ versus the vertex scatterers approach with constant
\(\boldsymbol{\sigma}^{v}\), as well as their consequences for quantum
chaos questions, have been explored in \cite{Car_ejde99} and
\cite{Ber_incol08}.  A definitive link has been established by
Harrison \cite{Har_rmp24} who showed that the secular equation
\(\det\left(\mathbb{I}_{2\left| \Bset \right|}-\mathbf{T}(k)
  \mathbf{S} \right)=0\), cf. \eqref{quantum_map}-\eqref{secular},
obtained by directly prescribing constant vertex scattering matrices
\(\boldsymbol{\sigma}^{v}\), matches the scular equation of a
\emph{Dirac operator} on the graph \cite{BulTre_jmp90,BolHar_jpa03},
describing a particle with zero mass and with no spin rotations at
vertices.

%%%%%%%%%%%%%%%%%%%%%%%%%%%%%%%%%%%%%%%%

\section{Quantum chaos on quantum graphs:
  eigenvalue statistics}
  \label{sec:spectrum}

Quantum graphs have become a paradigmatic system for quantum chaos after the seminal
papers of Kottos and Smilansky \cite{KotSmi_prl97,KotSmi_ap99} 
which showed that they are highly versatile systems that show
quantum signatures of chaos such as universal spectral fluctuations
(or transport statistics \cite{KotSmi_prl00,KotSmi_jpa03})  
as predicted by 
random-matrix theory \cite{QSoC} (see  also \cite{kieburg_2026, savin_fyodorov_2026} 
in this volume for random-matrix approaches to quantum chaos).
% \SG{See chapter RMT and Transport chapters in this volume}. 
Quantum graphs allowed to consider approaches based on sums 
over periodic orbits or trajectories 
numerically and analytically with much more ease than other quantum chaotic systems in 
the semiclassical approximation. For the latter finding a few 100 relevant classical trajectories 
and their relevant properties (action, stability, Maslov indices) is a major numerical effort 
and analytically out of reach -- for quantum graphs finding trajectories or periodic orbits
together with their amplitudes
just requires matrix multiplication. 
In many cases sums over relevant classes of trajectories can be 
performed analytically.
Apart from the analogy to semiclassical quantum chaos a further reason that made 
quantum graphs ideal paradigms for quantum chaos is their analogy to disordered system.
The effective disorder here is created by the quasi random phase factors that are collected 
when propagating along an edge and the subsequent scattering at vertices. 
In disordered systems effective field theories have been used to derive random-matrix 
behaviour and understand deviations from random-matrix theory in finite systems.
This is often referred to as the \textit{supersymmetry method} as the effective field theories
are based on a combination of commuting and anti-commuting (Grassmann) fields and 
a (broken) supersymmetry. We will not discuss the supersymmetry method here
and refer to the review 
\cite{altland2015review} which contains 
applications to general quantum chaotic systems including an overview
of approaches to quantum signatures of chaos in 
quantum graphs and a corresponding list of relevant references.
\\
 Here we will focus on a very short introduction
to some of the basic underlying ideas in the periodic-orbit approach
to study random-matrix behavior 
(focusing on spectral statistics). More detail and a more complete
list of relevant literature can be found in the 2006 review \cite{GnuSmi_ap06}.
After that  we will
 move on to give an overview of some current research directions for quantum graphs
in theoretical and experimental physics 
(some of which go beyond classical quantum chaos topics).
 We will come
back to further mathematical applications in Sec.~\ref{sec:mathematical-applications}.

\subsection{Spectral statistics  and universality for Quantum Graphs}

Universality in physics refers to a large number of phenomena where
large classes of physical systems (\textit{universality classes}) have
common (\textit{universal}) properties that are shared by all members
of the universality class irrespective of the system details. The laws
of thermodynamics in equilibrium systems is the most prominent but by
far not the only example of universality in physics.  Universality is
often connected to asymptotic limits (such as large system size for
thermodynamics). The Gaussian ensembles of random matrix theory define
such universality classes for many quantum signatures of chaos (in the
limit of large matrix dimension). If the origin of `chaos' is disorder
in the system then this universality can often be proven by an
ensemble average over the disorder (random-matrix theory itself may be
viewed as the most extreme case of this). For Hamiltonian systems in
the semiclassical limit random-matrix behaviour is the content of the
seminal Bohigas-Giannoni-Schmit conjecture
\cite{BohGiaSch_prl84}
(see also \cite{sieber_muller_2026} in this volume and \cite{QSoC}):
% \SG{see chapter by Martin and SEbastian}
 spectral correlations of classically
chaotic systems show the same spectral fluctuations as the Gaussian
ensembles of random-matrix theory (with three universality classes
related to the nature of time-reversal symmetry).  Universality is not
limited to spectral fluctuations and similar conjectures may be
formulated for other signatures, e.g. transport statistics in open
quantum systems (see \cite{novaes_2026,savin_fyodorov_2026} in this volume).
%\SG{refer to chapters by DIMA/Yan and the
%  semiclassical chapter by
%  Marcel, aslo 
%\cite{BerHarNov_jpa08,BerKui_pre12,BerKui_jmp13a,BerKui_jmp13b}.\\
For quantum graphs one may consider the equivalent conjecture in the
limit \(\left| \Eset\right| \to \infty\) of large graphs. This brings
about the question how to grow the size and whether the universality
holds for arbitrary large graphs and some fixed disorder in the edge
lengths (we will come back to explain this in more detail).  This
question has been answered in a negative way -- the first
counter-example that was analyzed in detail were star graphs with
Neumann-Kirchhoff conditions at the central vertex and any set of edge
lengths
\cite{BerKea_jpa99}.  Clear conditions that
lead to random-matrix behavior in the limit of large graphs are well
understood today (though short of mathematical proof, see previous review 
\cite{GnuSmi_ap06} 
for details).  Instead of giving a complete picture we will
focus here on the form factor of the quantum map of a closed graph
\begin{equation}
	K_n= \frac{1}{2 |\Bset|} \overline{\left| \mathrm{tr}\left( \mathbf{U}(k)^n\right) \right|^2}
\end{equation}
where the overline refers to the spectral average
\begin{equation}
 \overline{f(k)}= \lim_{K\to \infty} \frac{1}{K} \int_0^K f(k) dk .
\end{equation}
The form factor measures two-point spectral correlations for the eigenphases of the quantum
map (which is not the same but very closely related to two-point correlations in the wave 
number spectrum).
Let us write the \(n\)-th trace as
\begin{equation}
\mathrm{tr}\left(\mathbf{U}(k)^n\right)=
\sum_{p} \frac{n}{r_p} A_p e^{i k L_p}
\end{equation}
where the sum is over periodic orbits \(p\) of topological length \(n\).
Then we may perform the spectral average and obtain an exact
expression for the form factor as a double sum over periodic orbits
\begin{equation}
 K_n =  \frac{n^2}{2 \left|\Bset\right| }
 \sum_{p,p'}  \delta_{L_p L_p'}\frac{A_{p'}^* A_p}{r_{p'} r_p}
 \label{formfactor_po}
\end{equation}
where the Kronecker symbol \(\delta_{L_{p'} L_p}\) reduces the double sum to pairs
of periodic orbits that have the same length. Note that in general
(for sufficiently large \(n\)) there are many
different periodic orbits that have the same lengths. 

\subsubsection{Spectral averages and quasi-disorder}

Let us start with considering what is meant by disorder in the edge lengths and why 
this is important.
We will see that for quantum graphs
a reasonable notion for this is to require that edge lengths are
\textit{rationally independent} which means that there is no set of 
integers \((m_1, \dots, m_{|\Bset|})\in \mathbb{Z}^{|\Bset|}\)
such that \( \sum_{j=1}^{|\Bset|} |m_j| >0\) (they do not all vanish)
and \(\sum_{j=1}^{|\Bset|} m_j \ell_j =0\).
To see the implication of such an assumption
let us have a look at the periodic orbit expression \eqref{formfactor_po}
for the formfactor where the sum is over pairs of periodic orbits 
\(p\) and \(p'\) that share the same metric length
\(L_p= L_{p'}\). We may write the periodic orbit lengths
as \(L_p =\sum_{j=1}^{|\Bset|}  m_{p,j}\ell_j\)
and \(L_{p'}= \sum_{j=1}^{|\Bset|}  m_{p',j}\ell_j\)
where \(m{p,j}\) and \(m_{p',j}\) are non-negative integers that count how often
the \(j\)-th edge is visited by the periodic orbits.
For rationally independent edge lengths \(L_p= L_{p'}\) then implies 
\( m_{p,j}= m_{p',j}\) -- so both periodic orbits need to visit each edge the same number 
of times (but not necessarily in the same order or traverse edges in the same direction).
This connects the problem of evaluating the form factor to a combinatorial problem
of finding all periodic orbits that traverse the edges a given number of times.
Note that the remaining sum does not depend on the explicit choice of edge lengths. 
This already implies some level of universality:
the form factor (and other spectral correlation functions) for all graphs
that only differ by the choice of edge lengths are the same 
(as long as that choice is rationally independent).\\
In a wider setting one may show that the spectral average for a single graph
is equivalent to an ensemble average over disordered phases in a unitary quantum
walk on the directed edges of the graph (or a unitary network model). 
Let us come back to the observation 
that the quantum map \(\mathbf{U}(k)\) depends
on the wave number only in through phase factors \( \left(e^{i k \ell_1}, e^{ik \ell_2}, \dots
e^{ik \ell_{|\Bset|}}
\right)\) on the diagonal of the transport
matrix \( \mathbf{T}(k)\). 
This is one ingredient of the
Barra-Gaspard approach
 \cite{BarGas_jsp00} that we have discussed above 
 in \cref{sec:BG} where we discussed that the
the mapping \( k \to \mathbf{T}(k) \) can thus be viewed as
a dynamical system on the \(|\Bset|\)-torus \(\mathbb{T}^{|\Bset|}\) in a
 `time' \(k\). For rationally independent
edge lengths this dynamical system is \emph{ergodic} on the torus which means that
\begin{equation}
\overline{\mathcal{F}\left[\mathbf{T}(k) \right]}=
\frac{1}{(2\pi)^{|\Bset|}} \int_{\mathbb{T}^{|\Bset|}} d^{|\Bset|}\boldsymbol{\alpha}\ 
\mathcal{F}\left[\mathbf{T} (\boldsymbol{\alpha})\right]
\label{phase_disorder}
\end{equation}
for any (continuous)  function \(\mathcal{F} \). 
Here \(\mathbf{T} (\boldsymbol{\alpha}) \) is obtained from
\(\mathbf{T} (k) \) by replacing
\( e^{i k \ell_j} \mapsto e^{\alpha_j}\). The right-hand side of \eqref{phase_disorder}
is an average over \(|\Bset|\) independent phase factors.
The corresponding mapping \(\mathbf{U}(k) \mapsto 
\mathbf{T}(\boldsymbol{\alpha})\mathbf{S}_{\Gamma} \)
sends the quantum map of a graph at energy \(k^2\) to
a quantum walk on the directed edges with phase disorder such that
spectral statistics based on spectral averages for an individual quantum graph 
is  identical to spectral statistics based on an average over phase disorder in a 
unitary quantum walk.

\subsubsection{The spectral form factor in the diagonal approximation}

Let us continue to discuss the form factor with the assumption that the
edge lengths of the quantum graph are rationally independent.
The double sum \eqref{formfactor_po} over pairs \((p,p')\) of periodic orbits
of the same length \(L_p =L_{p'}\) for the form 
factor may be written 
as a sum of a diagonal part \(p=p'\) and a non-diagonal part \( p \neq p'\).
The basis of the so-called diagonal approximations 
in double sums of the type
\( \sum_{j,l=1}^N z_j^* z_l = \sum_{j=1}^N |z_j|^2
+ \sum_{j\neq l } z_j^* z_l\)
in general is the observation that
the diagonal part is a sum over over positive contributions with no cancellations
while the product of two amplitudes \(z_{j}^* z_l\) in general contains phase factors.
If the latter are sufficiently random the sum will be dominated by the diagonal contribution.
When applying this to quantum graphs we will see that time-reversal invariance
leads to further coherent  contributions from pairs of orbits where
one is a time-reversed version of the other -- these need to be included in the diagonal part.

The relevance of time reversal invariance in random-matrix theory
and quantum chaos has a long history which goes back to the three-fold way
of Wigner and Dyson (see \cite{kieburg_2026} in this volume or the 
textbook \cite{QSoC} and references therein). 
%\SG{Add crossreference to other chapters and QSoC}.
For quantum graphs it can be formulated in terms of a symmetry of the 
quantum map \(\mathbf{U}(k)\): a quantum graph is \emph{time-reversal invariant} if
there is diagonal unitary matrix \(\mathbf{G}\) such that
\begin{equation}
 	\mathbf{U}(k)= 
 	\mathbf{G} \mathbf{\Sigma} \mathbf{U}(k)^T \mathbf{\Sigma}\mathbf{G}^{-1}
\end{equation}
where \(\mathbf{\Sigma}\) is the the matrix that changes the direction of a 
directed bond \(\mathbf{\Sigma}_{\me_\sigma \me'_{\sigma'}}= 
\delta_{\me_\sigma \me'_{-\sigma'}}  \).
The matrix 
\( \mathbf{G}_{\me_\sigma \me'_{\sigma'}} = 
\delta_{\me_\sigma \me'_{\sigma'}}
 e^{i \gamma_{\me_\sigma}}
\) can be considered as a local gauge (a reference phase for each directed edge).
Quantum graphs with Kirchhoff-Neumann and the more general \(\delta\)-type conditions that 
we have introduced here are all time-reversal invariant with the choice
\( \mathbf{G}= \mathbf{T}(-k/2) \). In order to break time reversal symmetry in a 
quantum graph one either requires more general matching conditions or needs to add a
magnetic field. Here we will continue to focus on the time reversal invariant case.
For any trajectory \(\theta= \left( \me_{n\ \sigma_n}, \dots , 
\me_{0\, \sigma_0}\right)\)
or periodic orbit 
\(p=\overline{ \me_{n\ \sigma_n}, \dots , \me_{1\, \sigma_1}}\) we may introduce their 
time reversed
versions, denoted by \( \hat{\theta}= \left( \me_{0\ -\sigma_0}, \dots , 
\me_{n\, -\sigma_n}\right)\) 
and \(\hat{p}=\overline{\me_{1\ -\sigma_1}, \dots , \me_{n\, -\sigma_n}}  \) 
which traverse the 
same bonds in the opposite order and direction.
Time reversal invariance implies that for all periodic obits \(p\) its amplitudes is the same 
as the one of its time reversed partner \(A_p= A_{\hat{p}}\). When dividing the form factor into
diagonal and off-diagonal parts this is taken into account by writing
\begin{subequations}
\begin{align}
K_n = & K_n^{\mathrm{diag}}+ K_n^{\mathrm{off-diagonal}}\\
K_n^{\mathrm{diag}}=& 2
\frac{n^2}{2\left| \Bset \right|}\sum_{p}
\left(1- \frac{\delta_{p \hat{p}}}{2}\right)
 \frac{\left|A_p\right|^2}{r_p^2}\\
K_n^{\mathrm{off-diagonal}}=&
\frac{n^2}{2\left| \Bset \right|}\sum_{p,p': p'\neq p, p'\neq \hat{p}}
\delta_{L_p L_{p'}} \frac{A_{p'}^* A_p}{r_{p'}r_p}
\end{align}
\end{subequations} 
where the diagonal part contains the contributions \(p'=p\) and \(p'=\hat{p}\)
in \eqref{formfactor_po} and the off-diagonal part all the remaining pairs.
While one would expect an overall factor \(2\) in the diagonal
part for the two pairs of orbits that have the same amplitude one needs to
take account that there may be periodic orbits that are equal to their time reversed
partner \(p=\hat{p}\). Such \textit{self-retracing} orbits only give one contribution
and this is corrected with the factor \((1-\frac{\delta_{p\hat{p}}}{2})\). \\
The absolute square 
\(\left| A_p\right|^2 = \prod_{j=1}^n  
\mathbf{M}_{\me_{j\ \sigma_j}\ \me_{j-1\ \sigma_{j-1}} }\)  
of a quantum amplitude
along the periodic orbit  \(p\) can be viewed as a \emph{classical} amplitude
along this orbit. With the classical trace \(\mathrm{tr}\ \mathbf{M}^n =
\sum_p \frac{\left|A_p \right|^2}{r_p}
\)
and \(\tau \equiv \frac{n}{2 |B|}\)
one may rewrite the diagonal part as
\begin{equation}
K_n = 2 \tau \ \mathrm{tr}\ \mathbf{M}^n  - \tau \sum_{p} \delta_{p\hat{p}}
\frac{n \left|A_p \right|^2}{r_p^2} + 
2\tau \sum_{p}\frac{n(1-r_p)\left| A_p\right|^2}{r_p^2}\ .
\end{equation}
One expects that the diagonal part gives a dominant contribution when
\(n \ll 2 \left| \Bset \right|\) which leads to the
diagonal approximation \(K_n \approx K_n^{\mathrm{diag}}\).
We will not try to give any bounds on the off-diagonal part that would justify this approximation.
Instead we will, in the next section, show under what conditions the diagonal approximation
is consistent with universal random-matrix prediction  
\(K_n \sim 2 \tau + \mathcal{O}(\tau^2)\) for the form factor in the universality
class of the Gaussian orthogonal ensemble (time-reversal invariant systems).

%%%%%%%%%%%%
\subsubsection{Conditions for universality: Tanner's criterion}
\label{sec:tanner}

In numerical simulations one observes universal spectral
statistics in large extended graphs with many vertices.
In the following we will use the example  of a connected \(d\)-regular graph 
with connectivity \(d \ge 3\) (connectivity two is equivalent to a )
with \(|\Vset|\) vertices
and \(|\Bset| =\frac{d |\Vset|}{2}\) bonds as a guiding example.
We will consider the form factor \(K_n\) for large graphs \(|\Bset| \propto |\Vset| \to \infty\)
for short `time' \(\tau= \frac{n}{2|\Bset|}\ll 1\).
For large \(d\)-regular graphs the diagonal approximation can be justified for short times 
as pairs of orbits that would contribute to the off-diagonal part require that vertices are
visited more than once. On the same basis 
there is a negligible number of  orbits 
that are self-retracing or a repetition of a shorter orbit 
and we may focus on
\begin{equation}
	k_n \sim 2 \tau \ \mathrm{tr}\ \mathbf{M}^n
	= 2\tau \left(1 + \sum_{j=2}^{2B} \lambda_j^n \right) \ .
\end{equation}
Here \(\lambda_j\) for \(1\le j \le 2 |\Bset|\) are the eigenvalues of
the classical map with \(\lambda_1=1\) being the (unique!) unit
eigenvalue of the invariant (uniform) distribution, and
\(|\lambda_j| \le 1\) for all other eigenvalues.  For a fixed time the
terms
\( |\lambda_j^n|= e^{ \left(2 |\Bset |\, \log | \lambda_j|\right) \tau
}\).  If one considers an asymptotic limit \(B \to \infty\) (a
sequence of growing graphs) then the combined contributions of the
non-unit eigenvalues \(\lambda_j\) for \(2\le j\le 2 |\Bset|\)
vanishes pointwise at fixed \(\tau\) if they all satisfy
\( 1 - |\lambda_j| > c |\Bset|^{-\alpha}\) for some \(\alpha< 1\).  So
under this condition the short-time (\(\tau\ll 1\) form factor shows
universal random-matrix behaviour in leading order.  This is the
foundation of Tanner's conjecture \cite{Tan_jpa01} which extends the
statement to the full form factor at any value of \(\tau\) (with
pointwise convergence).  While the conjecture is hard to prove
rigorously a full derivation of the form factor at all times can be
obtained by the supersymmetry method for which we refer to
\cite{GnuAlt_prl04,GnuAlt_pre05,pluhar_2014}.  Within semiclassical
periodic-orbit theory one may include off-diagonal pairs that
correspond to Sieber-Richter pairs or higher-order encounters (see
\cite{sieber_muller_2026} in this volume and \cite{QSoC}).  The
corresponding expansion has been performed for quantum graphs with or
without time-reversal symmetry in
\cite{BerSchWhi_prl02,BerSchWhi_jpa03,Ber_incol06} and in the presence
of spin in \cite{BolHar_jpa03}.  In both the
supersymmetry and the periodic-orbit approaches it turns out that the
leading deviations from universality can be captured on the level of
the diagonal approximation.  A closer look at these deviation show
that Tanner's criterion does not exclude universality if the
conditions are not satisfied --- indeed the form factor is not a
self-averaging quantity and one would usually allow for an additional
average over time of the form
\begin{equation}
  \label{eq:K_time_averaged}
  K(\tau) = 
  \lim_{|\Bset| \to \infty} \frac{1}{2 \Delta n} 
  \sum_{ n :  \big|n-    2|\Bset| \tau   \big| < \Delta n} K_n,\quad
  \text{where}\quad \Delta n = c |\Bset|^{1-\xi}\quad \text{and}\quad
  \xi>0.
\end{equation}
This additional average over time suppresses contributions from eigenvalues 
that are not too close to unity even if their absolute value is close to unity.
For instance, any bi-partite graph (where vertices can be divided into two groups
and all bonds are between vertices in different groups) have at least one eigenvalue 
which is minus unity \(\lambda_2=-1\) which leads to strong oscillations in \(K_n\)
which are not present in the time-averaged version.
The detailed analysis shows that if there is any eigenvalues \(\lambda_j\) with \(j\ge 2\)
and \(\left| 1 - \lambda_j\right| < c/|\Bset|\) then the form factor will not follow random-matrix theory
and often show so-called \textit{intermediate} statistics (between 
quantum chaotic random-matrix statistics and Poissonian statistics known for integrable system).
If, however, \(\left| 1-\lambda_j\right|> c/\sqrt{|\Bset|}\) for all \(j\ge 2\) 
then the form factor and its Fourier transform (the two-point correlation function)
both converge pointwise to the universal random-matrix result.
If the distribution of eigenvalues is not covered by either of the above conditions then
both universal or non-universal limits are possible and one needs to consider the
collective limiting behavior of all eigenvalues. 
The spectral gaps  (see \Cref{sec:classicalmap})
\(\Delta \lambda= \mathrm{max}\left(\left| 1-\lambda_j\right|\right)_{j=2}^{2|\Bset|}\) or 
 \(\Delta \lambda^*\mathrm{max}\left(1-\left|\lambda_j\right|\right)_{j=2}^{2|\Bset|}\)
may be viewed as quantitative measures of chaos (analogue of classical Lyapunov exponents).
One may summarize the above findings by stating that in large graphs chaos needs to be 
sufficiently strong in the asymptotic limit \(|\Bset| \to \infty\).
In Hamiltonian systems the corresponding spectral gap in the 
Frobenius-Perron operator or the Lyapunov exponent are constant in the semiclassical
asymptotic limit \(\hbar \to 0\), in this sense chaos is always asymptotically strong
for classically chaotic systems.

In our  discussion of the diagonal approximation the contributions from
self-retracing orbits and repetitions of primitive orbits have not been treated correctly.
One can show that these contributions are safe to be neglected if the above conditions 
for universal behavior are satisfied. In the non-universal case these corrections are 
generally important.

Large well-connected graphs should generally be expected to follow
random-matrix theory as one can estimate the decay rates.  There are
however some exceptions to universal spectral statistics.  The most
well-known exception is star graphs with Neumann-Kirchhoff matching
conditions at the center \cite{BerKea_jpa99,BerBogKea_jpa01} where all
eigenvalues of the classical map are very close to one.  Further
deviations have been recently identified in \cite{BanExnGoeStr_jmp26}.
Other examples for non-universal behaviour (irrespective of whether
one uses Neumann-Kirchhoff matching conditions or more general
settings) are and graphs that grow in one dimension, tree graphs or
random \(d\)-regular graphs (which look a lot like tree graphs
locally). On the other hand \(d\)-regular graphs that are constructed
with many short cycles will usually show universal behaviour.  In a
generalized graph setting where the central scattering matrix is a
prescribed unitary matrix that leads to similar transition
probabilities for scattering in any edge (including backscattering)
star graphs do show universal behavior
\cite{BarGas_jsp00,Tan_jpa01,BerSchWhi_jpa03}.

%%%%%%%%%
\subsection{Wigner delay times}

When a wave pulse travels through a scattering region the scattered wave pulses 
usually have a certain delay compared to a free evolution.
When the wave pulses are very narrow in bandwitdh (i.e. one is close to
a monochromatic wave) then these delays can be described the so-called
Wigner-Smith delay matrix. For quantum graphs this is defined by
\begin{equation}
 \mathbf{Q}(k) = - i\ 
  \mathbf{S}_{\Gamma}(k)^{\dagger}\frac{d\mathbf{S}_{\Gamma}(k)}{dk}\ .
\end{equation}
In general quantum chaotic 
systems it is very difficult to describe delay statistics beyond the narrow
bandwidth limit. 
For quantum graphs Smilansky and Schanz developed a unified theory
that covers the whole range  from narrow to broad bandwidths 
\cite{smilansky2017delay, smilansky2018delay}.
The latter case is particularly relevant for experimental applications with short pulses.
In this linit they find strong quantum interference effects which lead to long delays with an
algebraic tail. We do not have the space to go into the details -- rather we would like to point
to very recent interesting experimental results \cite{giovannelli2025physical} 
where negative and imaginary time delays
in microwave networks were studied and modelled 
using a lossy quantum graph approach 
(where the quantum graph scattering matrix is non-unitary due to internal losses).

%%%%%%%%%%%%%%%%%%%%%%%%%%%%%
%%%%%%%%%%%%%%%%%%%%%%%%
\section{Quantum chaos on quantum graphs:
  wave function morphology}
\label{sec:wavefunctions}

\subsection{Quantum ergodicity, random wave 
model, and Anderson localization}

Quantum signature of chaos in wave functions can be discussed in
similar way.  The main difference is that apart from periodic obit
approaches that are in many ways similar to the ones for spectral
statistics above (and the supersymmetry approach) many quite general
results may be proofed rigorously.  When one speaks of a
\textit{(quantum) chaotic wave function} then, generally, this refers
to situations where, globally the wave function is delocalized
(spreads through all available space) and locally the wave function
cannot be distinguished statistically from a random superposition of
plane waves traveling in all available directions.  Both statements
can be formalized in general and for quantum graphs.  A full
discussion how to formalize these notions for quantum graphs is beyond
the scope of this manuscript. Instead we want to give a basic overview
of the concepts and point to some relevant literature.  In quantum
chaos one speaks of \textit{quantum ergodicity} if the probability
densities connected to (highly excited) energy eigenfunctions cannot
be distinguished from the classical invariant (ergodic) measure when
tested on scales that are much larger than the wavelength. For quantum
graphs quantum ergodicity is usually considered in the limit of large
graphs. Quantum ergodicity is typical for large well connected graphs
but fails, again, for star graphs
\cite{BerKeaWin_prl03,BerKeaWin_cmp04}.  In the physics literature it
is established that (similarly to the case of universal spectral
statistics) quantum ergodicity for large graphs holds if the
underlying classical map is sufficiently chaotic
\(\left| 1-\lambda_j\right|> c/|\Bset|\) for all \(j\ge 2\), while the
proerty does not hold for very weak chaos
\(\left| 1-\lambda_j\right|< c/|\Bset|^2\) for all \(j\ge 2\). For
intermediate scaling it depends on the details of the classical
spectrum.  Note that the critical exponents are different then the
ones for universal spectral statistics such that universal spectral
statistics implies quantum ergodicity but not vice versa.  We refer to
\cite{gnutzmann:2008,GnuKeaPio_ap10} for more details including a
discussion
of random wave behaviour on graphs.\\
Rigorous proofs of quantum ergodicity are available for various
classes of large quantum graphs, for instance graphs arising from
interval maps \cite{BerKeaSmi_cmp07}, graphs without back-scattering
\cite{BraWin_ahp16} and, more generally, for expander graphs
\cite{IngSabWin_jlms20,AnaIngSabWin_jmpa21}.

\subsection{Scars and topological resonances}
\label{sec:topological_resonances}

In some of the examples above we have already come across the phenomenon of
\textit{perfect scars} in quantum graphs.
In general \textit{scarring} in wave functions refers to a localization  
along rays or (classical) trajectories in the system. In generic Hamiltonian
systems such scars exist in both quantum chaotic and regular settings
-- their complete absence known as \textit{quantum unique ergodicity}
is a quite special property. While scars refer to some level of localization one
usually expects the wave function to leak everywhere in the available space.
Quantum graphs with \(\delta\)-type matching conditions allow for scarred energy 
eigenfunctions that
are completely localized on a proper closed subgraph \( \Gamma' \subset \Gamma\)
with \(\Gamma\setminus \Gamma' \neq \emptyset \)
such that the  wave function is identically zero 
\(\psi_{\me}(x_{\me})=0\)
for all  \(\me \in \Gamma\setminus \Gamma'\).
One may extend the definition to include open subgraphs but this definition
is better suited to avoid lots of case distinctions and covers the relevant
phenomena that we want to present here.
There is a very general way how to construct examples of such perfect scars.
For this consider an arbitrary quantum graph \(\Gamma'\) with an arbitrary
choice of edge lengths and \(\delta\)-type matching conditions at the vertices of \(\Gamma'\).
One implication of the time-reversal invariance of \(\Gamma'\) is that all energy eigenfunctions
of \(\Gamma'\) can be chosen real. 
Take any eigenfunction \(\psi_{\Gamma'}\) with energy \(E=k^2>0\) such that its nodal 
set is non-empty 
(see section~\ref{sec:nodal}). Then we can add additional edges to \(\Gamma'\)
at the nodal points of \(\psi_{\Gamma'}\). If one adds additional edges at a nodal
point on a vertex of \(\Gamma'\) one can keep the \(\delta\)-type matching condition
with an increased set of edges. If the nodal point is inside some edge the  one creates a 
new vertex at this point, attached some edges and chooses some
\(\delta\)-type matching condition at the new vertex. The new edges may be 
either leads or bonds. In the latter case one may leave them as dangling bonds or connect the
free ends to a further quantum graph structure.
In either case one constructs a larger graph \(\Gamma\) such that \(\Gamma'\)
is a proper subgraph (formally one first needs to replace \(\Gamma'\) by a basically 
equivalent graph with additional vertices and divided edges).
The wave function \(\psi_{\Gamma'}\) can be extended to a wave function \(\psi_\Gamma\) 
on the larger graph
\(\Gamma\) by setting \(\psi_{\Gamma}=\psi_{\Gamma'}\) inside the subgraph \(\Gamma'\)
and \(\psi_\Gamma=0 \) everywhere else. Note that the \(\delta\)-type 
matching conditions are then satisfied at all vertices on the boundary of \(\Gamma'\)
as a subgraph of \(\Gamma\) and \(\psi_\Gamma\) is an energy eigenfunction on
the larger quantum graph \(\Gamma\) which is also a perfect scar.
Note that this construction does not imply that any other eigenfunction on
\(\Gamma'\) may be extended to a perfect scar on the same larger graph \(\Gamma\)
(but of course one can start a new construction from its nodal points).

The construction above started from a small graph \(\Gamma'\). One can
also turn around and ask what kind of substructures in a given graph \(\Gamma\) 
with Kirchhoff-Neumann, or more generally \(\delta\)-type, matching conditions 
anywhere will lead
to perfect scars. The answer to this is stunningly general.
Let us consider the case where
 \( \Gamma' \subset \Gamma \)  (and \( \Gamma' \neq \Gamma\))
is either a cycle or a path %(see figure  \SG{add figure to explain})
between dangling bonds (with either Dirichlet or Neumann conditions at 
the dangling vertices of degree one at the ends).
Let us further assume that all edge lengths within \(\Gamma'\) 
are an integer multiples of some given length, that is for
all \(\me \in \Gamma'\) we have
\(\ell_{\me} = m_{\me} L\) for some \(L>0\)
and \(0<m_{\me} \in \mathbb{N}\).
Then there will be infinitely many energy eigenfunctions of \(\Gamma\)
with wave numbers  \(k_n =\frac{2\pi n}{L}\) 
for \(0<n \in \mathbb{N}\)  that are perfectly scarred on \(\Gamma'\).
The corresponding scarred wave function has the form 
\(\psi_{\me}(x_{\me}) = A \sin(k_n x_{\me})\) with nodal points on the 
vertices  (and a consistent choice of the arbitrary direction on each edge).
We have already seen this for a loop attached to a vertex 
(a cycle with a single edge). 
The construction for a path is analogous.
If the ratio of the total length of \(\Gamma'\) and \(L\) is even one can increase the set
of perfect scars to wave numbers  \(k_n =\frac{\pi n}{L}\) (dropping a factor two).
Note that any closed graph which has just one vertex of degree three or larger 
contains either a cycle or a path between dangling bonds as proper subgraphs.
So by an appropriate (rationally dependent) choice of edge lengths 
perfect scars appear in many quantum graph structures.
For open quantum graphs any perfect scar leads to a bound state in the continuum of
scattering states.

If there is a perfect scar inside \(\Gamma'\) then 
our expression \eqref{eq_scattering_matrix}
for the scattering matrix \(S_{\Gamma'}(k)\) has apparent poles
at the wave numbers \(k=k_n\) which seem to contradict the unitarity of \(S_{\Gamma'}(k)\).
Indeed poles can in  general be  removed. This is explicitly shown in
under the assumption that there are no degeneracies
\cite{lawrie2023closed}.

Perfect scars along cycles with more than one bond (or paths between
dangling bonds) are not stable under continuous change of the edge
lengths -- if the edge lengths are not exact integer multiples of some
common length then it is impossible to have nodal points on all
vertices along the cycle and the perfect scar leaks into the rest of
quantum graph.  This is often referred to as \textit{almost perfect
  scars} or \textit{topological scars}.  Now consider a cycle with
\(N\ge 2\) edges which are rationally independent. While the condition
for having nodal points on all vertices along the cycle can then never
be met exactly in the entire spectrum of the graph. However, given any
prescribed error (e.g. small distance of actual nodal points to the
vertices) one can find (infinitely many) wave numbers \(k\) with
smaller errors leading to (infinitely many) almost perfect scars of
any desired precision.  In the context of open graphs these almost
perfect scars have been named \emph{topological resonances}. A
resonance of an open graph \(\Gamma\) is a pole of the scattering
matrix \(S_{\Gamma}(k) \) for a complex values of the wave number
\(k\).  The positions of such poles generally change continuously when
the edge lengths are changed.  For the topological resonances such a
change of edge lengths can be used to move the position of the complex
wave number to the real axis (where the pole can be removed from the
scattering matrix and a bound state in the form of a perfect scar
appears).  Topological resonances and topological scars leave trace in
the tails of wave function statistics (or resonance width statistics)
that have different scaling then deviations expected from
random-matrix or disorder models or from semiclassical models for
billiards and related systems. The exponents of these deviations
depend on the girth of the graph (the smallest number of bonds along
any cycle) or one the number of bonds along paths between dangling
bonds \cite{gnutzmann:2013a}.  When topological resonances are excited
through in a scattering they are accompanied by strong amplification
of the incoming wave in the region of the corresponding subgraph and
(otherwise weak) nonlinearities can be amplified along
\cite{gnutzmann:2011}. For some rigorous results we refer to
\cite{BerKeaWin_cmp04,DavExnLip_jpa10,DavPus_apde11,BerWin_n18,CdVTru_ahp18}.

%%%%%%%%%%%%%%%%%%%%%%%%%%%%%%%%%%%%%%%%%%%
\subsection{Nodal statistics}
\label{sec:nodal}

Let $\Graph$ be a connected compact metric graph and let $H$ be a
self-adjoint Schr\"odinger operator on $\Graph$ (e.g.\
$H=-\frac{d^2}{dx^2}+V(x)$ on edges) with Neumann--Kirchhoff or
$\delta$-type vertex conditions.  Let $\lambda_n$ be the ordered
eigenvalues of $H$ (counting multiplicity), and let $\psi_n$ be a real
eigenfunction associated to $\lambda_n$.  Generically
\cite{Alo_jam24,Fri_ijm05,BerLiu_jmaa17}, $\psi_n$ corresponds to a
simple eigenvalue and does not vanish at any vertex.  We make this
assumption of genericity for the remainder of this section (breaking
this assumption can lead to sharp deviations \cite{HofKenMugPlu_ahp21}
for the ``universality'' expectations).

The \emph{nodal set} of $\psi_n$ is
\[
  Z(\psi_n):=\{x\in \Gamma \setminus \Vset^D:\ \psi_n(x)=0\},
\]
where $\Vset^D$ is the set of vertices with Dirichlet conditions (in
other words, the Dirichlet vertices are not included in the nodal
set).  When $\psi_n$ is generic, the nodal set consists of finitely
many points, and \emph{nodal domains} are obtained by cutting the
graph at these zeros.  In other words, a \emph{nodal domain} is a
connected component of $\Graph\setminus Z(\psi_n)$.

The \emph{nodal count} of $\psi_n$ is the number of points in the
nodal set $Z(\psi_n)$; we will denote it by $\phi(\psi_n)$.
We denote by
$\nu(\psi_n)$
the number of nodal domains of $\psi_n$; this quantity is called
the \emph{nodal domain count}.  Letting
\[
  \beta:=|E|-|V|+1
\]
to be the \term{first Betti number (cycle rank)} of the graph, we have
the following extension of the Sturm oscillation theorem and the
Courant nodal bound to the quantum graphs
\cite{BanBerSmi_ahp12,GnuSmiWeb_wrm04,Ber_cmp08},
\begin{equation}
  \label{eq:nodal_bounds_beta}
  n-\beta \ \le\ \nu(\psi_n)\ \le\ n,
  \qquad
  n-1 \leq \phi(\psi_n) \leq n-1 + \beta.
\end{equation}
Equivalently, the \emph{nodal deficiency} $\delta_n$ and the
\emph{nodal surplus} $\sigma_n$ defined by
\[
  \delta_n:=n-\nu(\psi_n),
  \qquad
   \sigma_n := \phi(\psi_n) - (n-1)
\]
both satisfy $0\le \delta_n, \sigma_n\le \beta$.  As a special case of
\eqref{eq:nodal_bounds_beta}, on a tree ($\beta=0$) one recovers the
Sturm-type nodal property $\nu(\psi_n)=n$ and $\phi(\psi_n)=n-1$.

Smilansky and his
collaborators
\cite{GnuSmiWeb_wrm04,BluGnuSmi_prl02,BanOreSmi_incoll08,KarSmi_jpa08}
have observed that the \emph{statistics} of the nodal count can be
used as a test for quantum chaos; the distribution of $\nu(\psi_n)$ is
drastically different from classically integrable to classically
chaotic systems.  In the chaotic case, including on graphs
\cite{GnuSmiWeb_wrm04}, this distribution, when appropriately
rescaled, was expected to converge to Gaussian.

The only case where this conjecture has been rigorously established
was on two families of quantum graphs: graphs with disjoint cycles
\cite{AloBanBer_cmp17}, mandarin (pumpkin) graphs, and flower graphs
with dangling edges \cite{AloBanBer_em22}, using the connection of the
nodal count to the stabiliy of the eigenvalues of the magnetic
Schr\"odinger operator on quantum graphs \cite{BerWey_ptrsa14}.
In the remainer of this section we review a
rigorous version of Smilansky's nodal universality conjecture and the
nodal-magnetic connection.

%%%%%%%%%%%%
\subsubsection{Nodal universality conjecture}

The nodal universality conjecture for quantum graphs was mentioned,
without specifics, in \cite{GnuSmiWeb_wrm04} and made rigorous in
\cite{AloBanBer_em22}.  In our exposition, we follow
\cite{AloBanBer_em22}.  

As shown in \eqref{eq:nodal_bounds_beta}, the \term{nodal surplus}
$\sigma_n := \phi(\psi_n) - (n-1)$ is an integer between $0$ and
$\beta$.  We define the probability to observe value $s$ as the nodal
surplus in a graph $\Gamma$ by
\begin{equation}
  \label{eq:prob_to_observe}
  \mathbb{P}(\sigma = s) :=
  \lim_{N\to\infty} \frac1N \Big| \left\{n \leq N: \sigma_n =
    s\right\} \Big|.
\end{equation}
Using Barra--Gaspard method \cite{BarGas_jsp00,BerWin_tams10}, see
Section~\ref{sec:BG}, one can
show (see \cite{AloBanBer_cmp17}) that this limit exists, defines a
valid distribution on $\{0,1,\ldots, \beta\}$ and, moreover, satisfies
$\mathbb{E} (\sigma) = \frac\beta2$.  The nodal universality
conjecture can be stated as follows.

%\begin{tcolorbox}
\begin{conjecture}
  \label{conj: Universality}
  Let
  $\left\{ \Gamma^{\left(\beta\right)}\right\} _{\beta\nearrow\infty}$
  be any sequence of standard graphs, with the first Betti numbers
  $\beta\to\infty$.  Choosing arbitrary rationally independent edge
  lengths for each $\Gamma^{\left(\beta\right)}$, let
  $\sigma^{\left(\beta\right)}$ denote its nodal surplus random
  variable. Then, the variance has linear growth
  $\var\left(\sigma^{(\beta)}\right) \sim \beta$ and in the limit of
  $\beta\rightarrow\infty$,
  \begin{equation}\label{eq: universality}
    \frac{\sigma^{\left(\beta\right)}-\frac{\beta}{2}}
    {\sqrt{\var\left(\sigma^{(\beta)}\right)}}
    \xrightarrow{\ \mathcal{D}\ }
    N(0,1),
  \end{equation}
  where the convergence is in distribution and $N(0,1)$ is the
  standard normal distribution.
\end{conjecture}
%\end{tcolorbox}

We stress that the conjecture claims uniform convergence over all
graphs with given $\beta$ and all choices of edge lengths.  Such a
demanding conjecture ought to be easier to refute, yet no known
counter-example exists (in addition to analytical results already
mentioned, \cite{AloBanBer_em22} contains a numerical study of the
``usual suspects'' families of graphs: quasi-one dimensional, square
lattice, regular, and complete).

%%%%%%%%%%%%
\subsubsection{Magnetic interpretation of the nodal surplus}

In the generic setting, $\sigma_n$ admits interpretations in terms of
the response of eigenvalues to magnetic fluxes through the cycles (a
``nodal--magnetic correspondence'' or ``Berkolaiko--Colin de Verd\`ere
theorem''), originally established on discrete graph (generalized)
Laplacians \cite{Ber_apde13,Col_apde13}.  The quantum graph version
was established in \cite{BerWey_ptrsa14}.

%\GB{Is it too much blowing my own horn?}

Magnetic Schr{\"o}dinger operators on quantum graphs are introduced in
\cite{KotSmi_ap99,GnuSmi_ap06,BerKuc_graphs} among other sources.
Since we did not introduce them in this article, we will proceed
directly to the description via the secular equation.  Let
$\alpha=\{\alpha_j\}$ be real values (``magnetic potentials'')
associated with non-directed edges $\me_e\in\Eset$.  We define the
diagonal $\Eset_\pm$-indexed matrix $\mathbf{T}(k, \alpha)$ by
\begin{equation}
  \label{eq:Talpha}
  \mathbf{T}(k, \alpha)_{\me_{j,s}, \me_{j,s}} =
  \begin{cases}
    e^{i(k\ell_j + \alpha_j)} & \text{if } s=+,\\
    e^{i(k\ell_j - \alpha_j)} & \text{if } s=-.
  \end{cases}
\end{equation}
When $\alpha \equiv 0$, we recover the matrix $\mathbf{T}(k)$ from \eqref{eq:Tdef}.
For a graph with Neumann--Kirchhoff conditions and magnetic potentials
$\alpha$, the \term{magnetic secular equation} is 
\begin{equation}
  \label{eq:mag_secular}
  \det(1-\mathbf{T}(k, \alpha)\mathbf{S})=0.
\end{equation}

\begin{thm}\cite{BerWey_ptrsa14}\label{thm: nodal magnetic}
  Let $\lambda_n = k_{n}^2$ be a simple eigenvalue of the graph
  $\Gamma$ with an eigenfunction that does not vanish on vertices.
  Let $k_{n}(\alpha)$ be a solution of \eqref{eq:mag_secular}
  extending $k_n$ (i.e. $k_n(0)=k_n$).

  Then $\alpha=0$ is a critical point of $k_{n}(\alpha)$ with the Morse index $\sigma_n$.
\end{thm}

In the theorem, the \term{Morse index} is the number of negative eigenvalues
of the Hessian (the matrix of second derivatives)
\begin{equation}
  \label{eq:Hess_def}
  \Hess k_{n}(0) = \left( \frac{\partial^2 k_n}{\partial \alpha_i
      \partial \alpha_j}(0) \right)_{i,j=1}{|\Eset|}.
\end{equation}
In a small departure from \cite[Thm.~2.1]{BerWey_ptrsa14}, we introduced
$E-\beta$ ``extra'' parameters $\alpha$ (which can be removed by the
magnetic gauge invariance), leading to the Hessian having a kernel of
dimension $E-\beta$.

%\GB{Maybe: mention persistent currents?}
%\SG{I think we will always find new things that could be added and are just as interesting as 
%stuff that we actually did mention. I think we call it a day (and apply that to inside -outside 
%unless you have added that aleady in the parts below that I have not yet read) as far as the 
%first arxiv version is concerned. We can still add something before the end of the month.}

%%%%%%%%%%%%%

%%%%%%%%%%%%%%%%%%%%%%%%%%%%%%%%%%%%%%%%%%%%%%%%%%%%%
\section{Further applications}
\label{sec:applications}

\subsection{Fourier quasicrystals}
\label{sec:FQ}

%\SG{Maybe formulate the following in physics notation, i.e. density 'functions' rather than 
%'measures' and writing rename 'atomic measure' to a delta comb or similar?}
A \emph{Fourier quasicrystal} (FQ) is discrete set $K\subseteq\R^{d}$
such that the Fourier transform of the Dirac comb $\mu =
\sum_{k\in K}\delta_k$ supported on $K$ is also a Dirac comb:
\begin{equation}
  \label{eq:FTmu}
  \hat\mu = \sum_{\xi\in\hK} c_{\xi}\delta_\xi,
\end{equation}
for some discrete set $\hK\subseteq\R^{d}$ (called the \term{spectrum
  of the quasicrystal}) and some complex numbers $(c_{\xi})_{\xi\in\hK}$.
In addition, a technical condition is usually imposed: $\mu$,
$\hat\mu$, $|\mu|$ and $|\hat\mu|$ are tempered distributions.

The first condition can be formulated as follows: for every Schwartz
function $f\in\mathcal{S}(\R^{d})$, the equality
\begin{equation}
  \label{eq:FTtest}
  \sum_{x\in K}\hat{f}(x) = \sum_{\xi\in \hK}c_{\xi}f(\xi)
\end{equation}
holds (and the sums are finite).  The technical condition is equivalent
to a polynomial growth bound on the sets $K$, $\hK$ and the
coefficients $c_\xi$, which we will not discuss here.

Equation~\eqref{eq:FTtest} is reminiscent of the Poisson summation
formula, and a periodic lattice satisfies the definition of an FQ.
One is usually interested in \emph{non-periodic} FQs.  The classical
method of their construction is \emph{cut-and-project} technique
\cite{LevSte_prb86}, but it yields quasicrystals with an unphysical
property: the atoms get arbitrarily close to each other.

A harder challenge is to find a Fourier quasicrystals $K$ which is a
\term{Delone set}: there are real numbers $r$ and $R$ such that any
two points are separated by at least $r>0$ and any
interval of length $R$ contains at least one point.

Lagarias \cite{Lag_crm00} conjectured that if an FQ has Delone
support $K$ and Delone spectrum $\hK$, then it is periodic; 15 years later
Lev and Olevskii \cite{LevOle_im15} proved this conjecture, but not a
single Delone non-periodic quasicrystal was known at the time.

An attentive reader would observe that we have seen an equation
similar to \eqref{eq:FTtest} in Section~\ref{sec:trace_formula},
namely the Roth--Kottos--Smilansky trace formula,
equation~\eqref{eq:full_trace_formula}.  This connection, namely that
the $k$ spectrum of a quantum graph is a Fourier quasicrystal, by
virtue of the trace formula
\cite{KotSmi_prl97,KotSmi_ap99,Rot_incol84} was made by Kurasov and
Sarnak \cite{KurSar_jmp20}, who then went on to produce the first
example of a Delone quasicrystal.

The difficulty in producing a Delone FQ is that most graphs have
eigenvalue pairs with arbitrarily small gaps
\cite{BarGas_jsp00,CdV_ahp15,Alo_jam24}.  However, Kurasov and Sarnak
noticed that the symmetric part of the tadpole graph spectrum, see
equation \eqref{eq:sec_tadpole_sym}, possesses a lower bound on the
nearest-neighbor spacing.

\begin{figure}
  \centering
  \includegraphics[scale=0.25]{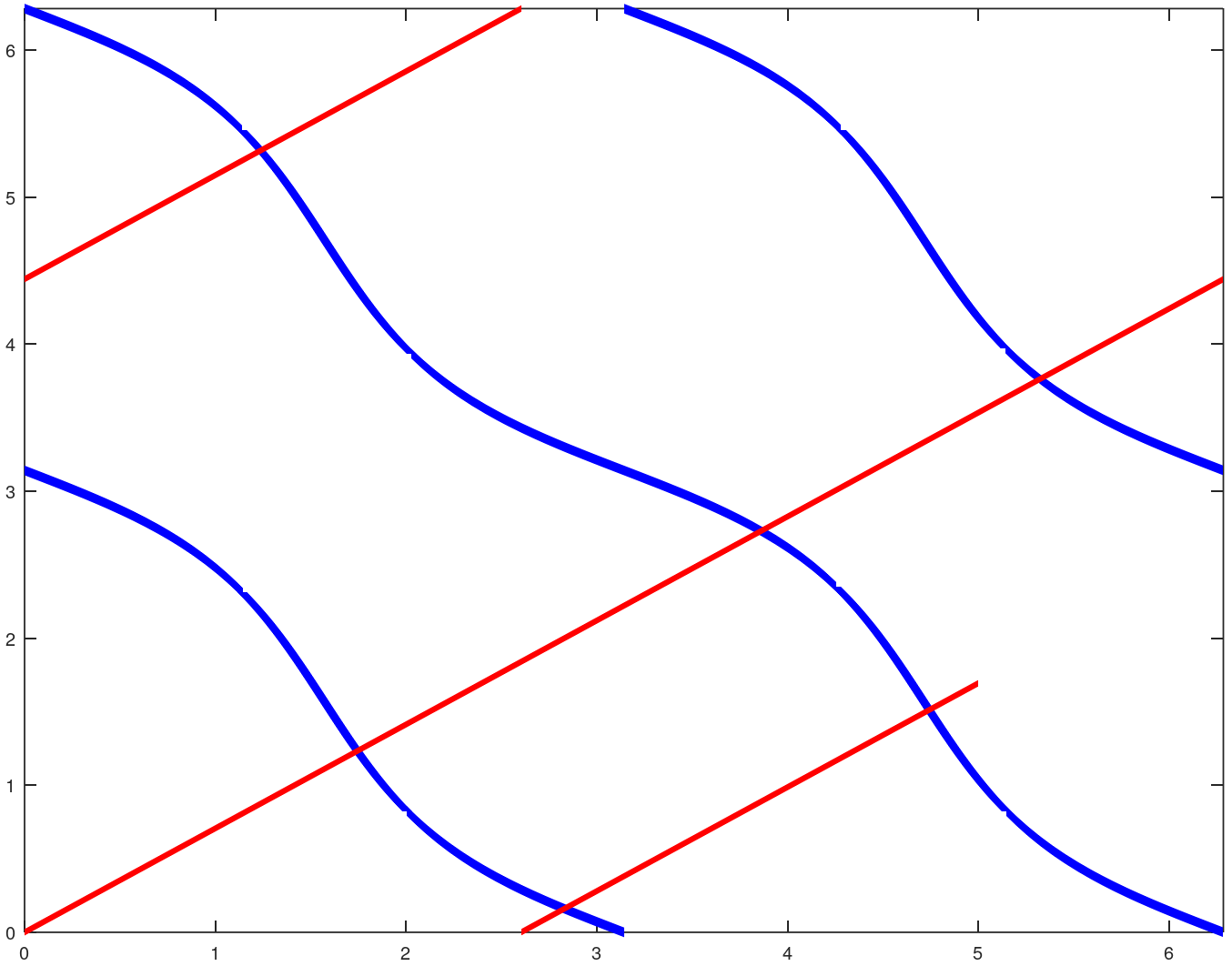}
  \caption{The solution set of equation~\eqref{eq:sec_on_torus}, shown
    in blue, intersected by a linear flow, shown in red, with
    $\ell_1=1$ and $\ell_2=\sqrt{2}$.}
  \label{fig:KS_secular}
\end{figure}

The latter observation can be verified using the Barra--Gaspard
technique, see Section~\ref{sec:BG}.  We view the roots $k$ of
\eqref{eq:sec_tadpole_sym} as the intersection times of the flow
$k \mapsto \Big( \ell_1k, \frac{\ell_2}{2} k\Big)$ with the
solution set of the equation
\begin{equation}
  \label{eq:sec_on_torus}
  \sin(\xi_1-\xi_2) - 3\sin(\xi_1+\xi_2) = 0,
  \qquad
  (\xi_1,\xi_2) \in (0, 2\pi]^2,
\end{equation}
on the torus, see Figure~\ref{fig:KS_secular}.
The solution set has no self-intersections and therefore no small gaps.

\subsection{Modelling metamaterials using quantum graphs}
\label{sec:metamaterials}

Metamaterials are composite physical compounds with an engineered structure
on scales that are much larger than the atomic structure -- reaching from nano- to 
macro-scales. They can be used to manipulate (accoustic, electromagnetic, seismic, 
or quantum) wave propagation through these material in ways that do not occur in
nature. Examples are negative refractive indices, perfect lensing and cloaking.
The versatility of quantum graph models can be exploited in this context and, though not
directly related to quantum chaos, we would like to mention some recent work by Lawrie, Tanner
\textit{et al}
\cite{lawrie2022quantum,lawrie2024application,lawrie2025nondiffracting} where
negative refraction and angular filtering in quantum graph models for metamaterials 
was demonstrated.

\begin{figure}
\includegraphics[width=0.48\textwidth]{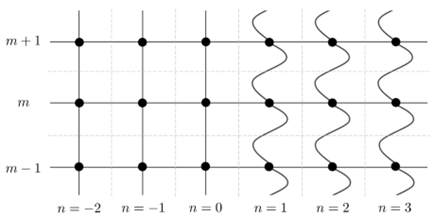}
\hfill
\includegraphics[width=0.42\textwidth]{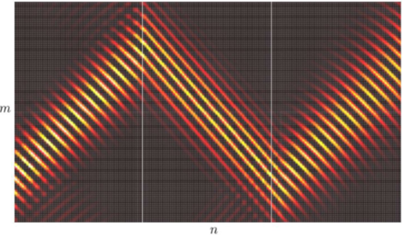}
\caption{\label{fig:metamaterial-qg}
Left: 
[Reprinted from \cite{lawrie2022quantum} with permission of the authors]
Boundary between two quantum graph metamaterials. On the left side of the 
boundary vertices and edges form a regular 2D square lattice with the same edge lengths 
in horizontal and vertical directions. On the right of the boundary the underlying lattice is the 
same but the vertical edge lengths are different from the horizontal.\\
Right:
[Reprinted from \cite{lawrie2022quantum} with permission of the authors]
Wave refraction of a stationary wave 
from two boundaries between quantum graph metamaterials.
The picture clearly demonstrates a negative refraction index as the transmitted wave changes its
vertical direction at each boundary.
}
\end{figure}

Negative refraction was first simulated \cite{lawrie2022quantum}
and later demonstrated experimentally using acoustic waveguides
\cite{lawrie2024application}
in a simple setting where metamaterials were designed
based on a 2D square lattice of edges and vertices.
Different materials may be created by either manipulating matching
conditions or the choice of edge lengths. Fig.~\ref{fig:metamaterial-qg}
shows the actual setting (left panel) that was used in the simulation
a negative refraction index (right panel).

\begin{figure}
\includegraphics[width=0.48\textwidth]{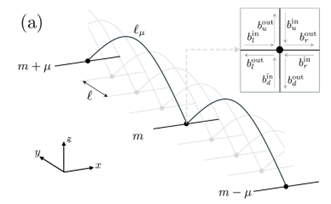}
\hfill
\includegraphics[width=0.42\textwidth]{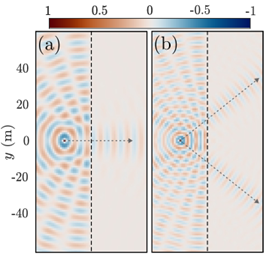}
\caption{\label{fig:filtering-qg}
Left: 
[Reprinted from \cite{lawrie2025nondiffracting}]
Model of a boundary between two identical quantum graph lattices with rewired
edges connecting vertices several lattice spacings apart.
\\
Right:
[Reprinted from \cite{lawrie2025nondiffracting} with permission of the authors]
Demonstration of angular filtering. A stationary wave is excited at one point left of the boundary.
Depending on the choice of parameters different directions are transmitted.
}
\end{figure}

Angular filtering was demonstrated
\cite{lawrie2025nondiffracting}  by creating a boundary between
two (identical) 2D regular square  quantum graph lattices by rewiring the
the edges along the boundary as shown in the left panel of 
Fig.~\ref{fig:filtering-qg} using longer edges that connect vertices that are several
lattice spacings apart -- the right panel shows a simulation 
of two a stationary wave that is excited at one point on the left panel
and hits a boundary such that only certain angular directions are transmitted
(where the angle depends on the parameters that can be manipulated).

%%%%%%%%%%%%%%%%%%%%%%%%
%%%%%%%%%%%%%%%%%%%%%%%%%%%%%
\section{Further mathematical techniques}
\label{sec:mathematical-applications}

%%%%%%%%%

%%%%%%%%%
\subsection{The Dirichlet-to-Neumann (DtN) map}
\label{sec:dtn}

An alternative to the scattering approach, which focuses on wave
amplitudes on directed edges, is the approach of the
Dirichlet-to-Neumann (DtN) map (see Section 3.5 in
\cite{BerKuc_graphs} as well as \cite{KucZha_jmp19}), which focuses on
the wave function values at the vertices.

A subset $\partial \Vset$ of the vertex set $\Vset$ is declared to be the
\term{boundary} of the graph.  Without loss of generality we take
$\partial \Vset = \{\mv_1, \ldots, \mv_N\}$ with \(N= \left|\partial \Vset \right| \).
  The
\emph{Dirichlet-to-Neumann map} $\Lambda(k)$ relates the vector of
boundary vertex values
$\boldsymbol{\Psi} = (\psi(\mv_1), \dots, \psi(\mv_{N}))^T$ to
the vector of derivatives of the solution to the Schrödinger equation
$-\psi'' = k^2\psi$, also at the boundary.  The vertex values are the
``Dirichlet data'' and the derivatives are the ``Neumann data''.  In
keeping with our convention, the derivatives are taken pointing into
the graph; if the boundary vertex has degree larger than 1, the
Neumann data is the sum of the derivatives on the individual edges.
One can imagine an experimenter attaching a lead to each boundary
vertex and reporting the current flowing along the lead into the
vertex, once the Neumann-Kirchhoff vertex condition has been imposed.
Note that the spectral parameter $k$ is allowed to be complex (which
helps to avoid singularities).

\subsubsection{Single edge DtN map}

On any single edge $\me \in \Eset$ of length $\ell_\me$ connecting
vertices $\mv_i$ and $\mv_j$, the Schrödinger equation
$-\psi'' = k^2\psi$ admits a general solution. If we prescribe the
values of the wave function at the endpoints (Dirichlet boundary
conditions) to be $\psi(\mv_i)$ and $\psi(\mv_j)$, the solution on the
edge can be written explicitly as:
\begin{equation}
  \label{eq:edge_solution_dtn}
  \psi_\me(x_\me) = \frac{1}{\sin(k\ell_\me)}
  \left( \psi(\mv_i) \sin(k(\ell_\me - x_\me)) + \psi(\mv_j) \sin(k x_\me) \right),
\end{equation}
where $x_\me$ is the coordinate along the edge increasing from $\mv_i$ ($x_\me=0$) to $\mv_j$ ($x_\me=\ell_\me$). We assume here that $k$ is not a Dirichlet eigenvalue of the edge, i.e., $\sin(k\ell_\me) \neq 0$.

A standard computation yields
\begin{equation}
  \label{eq:edgeDtN}
  \begin{pmatrix}
    \psi'_\me(0) \\
    -\psi'_\me(\ell_\me)
  \end{pmatrix}
  =
  \begin{pmatrix}
    -k\cot(k\ell_\me) & \frac{k}{\sin(k\ell_\me)} \\
    \frac{k}{\sin(k\ell_\me)} & -k\cot(k\ell_\me)
  \end{pmatrix}
  \begin{pmatrix}
    \psi(\mv_i) \\
    \psi(\mv_j)
  \end{pmatrix}
  =:
  \begin{pmatrix}
    A_\me(k) & B_\me(k) \\
    B_\me(k) & A_\me(k)
  \end{pmatrix}
  \begin{pmatrix}
    \psi(\mv_i) \\
    \psi(\mv_j)
  \end{pmatrix}  
\end{equation}
The matrix in \eqref{eq:edgeDtN} is the DtN map of the edge $\me$.

A collection of single edge DtN maps arises in the Behrndt--Luger formula for the
number of negative eigenvalues of a quantum graphs, see
\cite{BehLug_jpa10} or \cite[Sec.~3.5.5]{BerKuc_graphs}.

%%%%%%%%%%%
\subsubsection{All vertex DtN map}

The next simplest case is when every vertex in a graph is a boundary
vertex, $\partial \Vset = \Vset$ (so \(N= |\Vset|\)).  The Neumann data at a vertex $\mv$ is
obtained by summing up the contributions
from \eqref{eq:edgeDtN} over the edges incident to $\mv$.  The result
is the $N \times N$ matrix $\Lambda(k;\Vset)$ with entries
\begin{equation}
  \label{eq:dtn_matrix_elements}
  \Lambda_{\mv \mv'} = 
  \begin{cases} 
    \displaystyle \sum_{\me \sim \mv} A_\me(k) & \text{if } \mv = \mv', \\
    \displaystyle \sum_{\me \in \Eset_{\mv \mv'}} B_\me(k)
    & \text{if } \mv \sim \mv', \\
    0 & \text{otherwise}.
  \end{cases}
\end{equation}
Here $\Eset_{\mv \mv'}$ denotes the set of edges connecting $\mv$ and
$\mv'$, while we assumed the graph has no looping edges; the formula can be
adjusted to accommodate loops as well.

The matrix \eqref{eq:dtn_matrix_elements} allows us to give an
alternative form for the secular equation.  Recall the $\delta$-type
conditions from Eq.~\eqref{eq:cond_current}, which, in terms of the
DtN map, become
$(\Lambda(k;\Vset)\boldsymbol{\Psi})_\mv = \alpha_\mv \psi(\mv)$.
Collecting these for all vertices, the condition for the existence of
a non-trivial solution (an eigenstate) is:
\begin{equation}
    \label{eq:secular_dtn}
    \det \left( \Lambda(k; \Vset) - \Theta \right) = 0,
\end{equation}
where
$\Theta = \text{diag}(\alpha_{\mv_1}, \dots,
\alpha_{\mv_{|\Vset|}})$. For the standard Neumann-Kirchhoff
conditions, this simplifies to $\det(\Lambda(k; \Vset)) = 0$.

\begin{remark}
  \label{rem:masking}
  This formulation is often advantageous for numerical
  computations as the dimension of $\Lambda(k; \Vset)$ is equal to the
  number of vertices $|\Vset|$, which is generally smaller than the
  $2|\Eset|$ dimension of the scattering matrix $\mathbf{U}(k)$.
  Unfortunately, the poles of $A_\me(k)$ and $B_\me(k)$ occurring at
  Dirichlet eigenvalues of individual edges may mask some of the
  eigenvalues \cite{KucZha_jmp19}.

  There is a more robust variant of this method, looking at a
  \term{graph} of the DtN map, the so-called \term{Cauchy data space}.
  Rather than have poles at Dirichlet eigenvalues, the Cauchy data
  space becomes ``vertical'', which is not a singularity.  In fact,
  the Cauchy data space rotates smoothly as a function of the
  parameter $k$ \cite[Theorem~4.1]{BerCoxLatSuk_jst25} and the
  eigenvalues of the graph can be detected as intersection with
  another subspace encoding the vertex conditions.  This viewpoint
  leads naturally to using Maslov index for counting eigenvalues
  \cite{BerCoxLatSuk_jst25,HowSuk_jde16,LatSukSuk_jam18,LatSuk_cm20}
  as well as nodal domains \cite{BanProSof_prep25}.
\end{remark}

\begin{example}[Tadpole graph with a $\delta$ condition]
  \label{eq:delta_tadpole}
  We have not yet considered an example of a graph with
  $\alpha_\mv\neq 0, \infty$.  Equation~\eqref{eq:secular_dtn} is
  perfectly adapted to treating such graphs!

  Consider the quantum tadpole graph from
  Figure~\ref{fig:star}(left).  We impose a \(\delta\)-type condition
  at the connection vertex \(\mv_2\) with parameter \(\alpha\), and
  Neumann-Kirchhoff at the degree-one vertex \(\mv_1\).  Because
  equation~\eqref{eq:dtn_matrix_elements} assumed the graph has no
  loops, we will break the loop by introducing a ``dummy'' vertex
  \(\mv_3\) in the middle of it.  The condition at \(\mv_3\)
  is Neumann--Kirchhoff.

  Define the edge DtN coefficients for \(\me_1\) and for the half-loop
  edges (which are identical and therefore use the same coefficients
  $A_2$ and $B_2$):
  \begin{equation}
    \label{eq:ABtadpole}
    A_1=-k\cot(k\ell_1),\quad B_1=\frac{k}{\sin(k\ell_1)},\qquad
    A_2=-k\cot\Bigl(k\frac{\ell_2}{2}\Bigr),\quad
    B_2=\frac{k}{\sin\bigl(k\frac{\ell_2}{2}\bigr)}.
  \end{equation}
  Assembling according to \eqref{eq:dtn_matrix_elements} (remembering the two
  parallel edges), in the ordered basis \((\mv_1,\mv_2,\mv_3)\) we obtain
  the all-vertex DtN map
  \[
    \Lambda(k;\Vset)=
    \begin{pmatrix}
      A_1 & B_1 & 0 \\
      B_1 & A_1+2A_2 & 2B_2 \\
      0 & 2B_2 & 2A_2
    \end{pmatrix}.
  \]
  The \(\delta\)-parameters are $\Theta=\diag(0,\alpha,0)$.
  Evaluating the determinant in \eqref{eq:secular_dtn} and using
  $A_j^2-B_j^2=-k^2$ yields the condition
  \[
    A_2k^2 + 2A_1k^2 + \alpha A_1A_2 = 0.
  \]
  Substituting \eqref{eq:ABtadpole} and dividing by \(k^2\) (for
  \(k\neq 0\)), we obtain the secular equation
  \[
    k\cot\Bigl(k\frac{\ell_2}{2}\Bigr) + 2k\cot(k\ell_1)
    - \alpha \cot(k\ell_1)\cot\Bigl(k\frac{\ell_2}{2}\Bigr) =0,
  \]
  or its pole-free version
  \begin{equation}
    \label{eq:delta_tadpole_sec}
    k\cos\Bigl(k\frac{\ell_2}{2}\Bigr)\sin(k\ell_1)
    +2k\cos(k\ell_1)\sin\Bigl(k\frac{\ell_2}{2}\Bigr)
    -\alpha \cos(k\ell_1)\cos\Bigl(k\frac{\ell_2}{2}\Bigr)
    =0.    
  \end{equation}

  Two remarks are in order.  First, setting $\alpha=0$ we recover the
  secular equation for the NK tadpole as derived in
  \eqref{eq:sec_tadpole_sym}.  Second, equation
  \eqref{eq:delta_tadpole_sec} completely misses the loop states, as
  cautioned in Remark~\ref{rem:masking}, because they correspond to
  the Dirichlet spectrum of the loop edge(s).

  Finally, we mention that the introduction of the dummy vertex is not
  a necessary step.  With only two vertices $\mv_1$ and $\mv_2$, the
  DtN map description in \eqref{eq:dtn_matrix_elements} can be
  adjusted in a natural way to yield the secular equation
  \begin{equation}
    \label{eq:tadpole_secular_det}
    0=\det\bigl(\Lambda(k;\Vset)-\Theta\bigr)
    =
    \det\begin{pmatrix}
      A_1 & B_1\\
      B_1 & A_1 + 2A_2 + 2B_2 - \alpha
    \end{pmatrix},
    \qquad \text{with }
    A_2=-k\cot(k\ell_2),\quad B_2=\frac{k}{\sin(k\ell_2)},
  \end{equation}
  and $A_1$, $B_1$ as in \eqref{eq:ABtadpole}.  The result of
  evaluating \eqref{eq:tadpole_secular_det} is identical to
  \eqref{eq:delta_tadpole_sec}. 
\end{example}

%%%%%%%%%
\subsubsection{The graph DtN map}

We now consider the case when $\partial \Vset$ is a proper subset of $\Vset$.
The idea is to start with \eqref{eq:dtn_matrix_elements} and to
enforce vertex conditions at the interior vertices $\Iset := \Vset
\setminus \partial\Vset$.  The values of $\psi$ at the interior vertices need
to be eliminated, which results in a Schur complement.

More precisely, we partition $\Lambda(k; \Vset)$,
$\boldsymbol{\Psi}_{\Vset}$ and the Neumann data
$\boldsymbol{\Psi'}_{\Vset}$ into blocks according to the disjoint
union $\Vset = \partial \Vset \sqcup \Iset$:
\begin{equation}
  \label{eq:dtn_block_system}
  \begin{pmatrix}
    \boldsymbol{\Psi'}_\Bset \\
    \boldsymbol{\Psi'}_\Iset
  \end{pmatrix}
  =
  \begin{pmatrix}
    \Lambda_{\Bset\Bset} & \Lambda_{\Bset\Iset} \\
    \Lambda_{\Iset\Bset} & \Lambda_{\Iset\Iset}
  \end{pmatrix}
  \begin{pmatrix}
    \boldsymbol{\Psi}_\Bset \\
    \boldsymbol{\Psi}_\Iset
  \end{pmatrix}.
\end{equation}
The Dirichlet-to-Neumann map $\Lambda(k;\partial\Vset)$ is defined by the
relation $\mathbf{F}_{\partial \Vset} = \Lambda(k;\partial\Vset) 
\boldsymbol{\Psi}_{\partial \Vset}$
for a solution that satisfies vertex conditions at all interior
vertices.  At this point we assume that these conditions are either
Dirichlet or Neumann-Kirchhoff (simple adjustments can be made for
$\delta$-type conditions, by subtracting the $\Theta$ matrix as in
\eqref{eq:secular_dtn}).

First, the rows and columns corresponding to vertices in $\Iset$ with
Dirichlet conditions are \emph{removed} from the matrix
$\Lambda(k; \Vset)$.  Next, the variables $\boldsymbol{\Psi}_\Iset$
are eliminated from \eqref{eq:dtn_block_system} using the
Neumann--Kirchhoff condition $\boldsymbol{\Psi'}_\Iset = 0$.  Assuming
the Dirichlet vertices have already been removed from
\eqref{eq:dtn_block_system}, this results in the Schur complement
\begin{equation}
  \label{eq:dtn_schur_complement}
  \Lambda(k; \partial\Vset)
  = \Lambda_{\partial \Vset\ \partial\Vset}
  - \Lambda_{\partial \Vset\ \Iset} \Lambda_{\Iset\ \Iset}^{-1}
   \Lambda_{\Iset\ \partial \Vset}.
\end{equation}
The invertibility of $\Lambda_{\Iset\ \Iset}$ fails if and only if $k^2$
is an eigenvalue for the graph obtained by imposing Dirichlet
conditions at all vertices in $\Bset$ (this is a well-known issue with the
DtN maps that may be avoided by working with the so-called \emph{Cauchy
  data spaces}, see \cite[Sec.~7.5]{BerCoxLatSuk_jst25} for an example).

We also note that our derivation of the Dirichlet-to-Neumann map is
largely parallel to the derivation of the scattering matrix in
\Cref{sec:scatter_states}.  We now illustrate formula
\eqref{eq:dtn_schur_complement} with examples.

\begin{example}[Star Graph]
  \label{ex:star_graph_dtn}
  Consider a star graph with four edges $\me_1,\ldots,\me_4$ of
  lengths $\ell_1, \dots, \ell_4$ connected to a central vertex $v$,
  see Figure~\ref{fig:star_DtN}. We designate the endpoints of edges
  $\me_1$ and $\me_2$ as the boundary $\Bset$ and impose Dirichlet
  conditions at the endpoints of $\me_3$ and $\me_4$.

  \begin{figure}
    \centering
    \begin{tikzpicture}[scale=1]
      % Central vertex
      \coordinate (V) at (0,0);
      \draw[fill] (V) circle (1.5pt) node[below=0.1cm] {$v$};
      
      % Leaves
      \coordinate (L1) at (-2,1);
      \coordinate (L2) at (-2,-1);
      \coordinate (L3) at (2,1);
      \coordinate (L4) at (2,-1);
      
      % Edges
      \draw[->,thick] (V) -- (L1) node[midway, above] {$\ell_1$};
      \draw[->,thick] (V) -- (L2) node[midway, below] {$\ell_2$};
      \draw[thick] (V) -- (L3) node[midway, above] {$\ell_3$};
      \draw[thick] (V) -- (L4) node[midway, below] {$\ell_4$};
      
      % Leaf nodes
      \draw[fill=white] (L3) circle (1.5pt);
      \draw[fill=white] (L4) circle (1.5pt);
    \end{tikzpicture}
    \caption{Star graph with a 2-vertex boundary.}
    \label{fig:star_DtN}
  \end{figure}
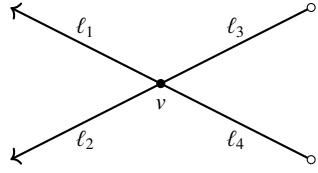

  As before, we denote
  \[
    A_j(k) := -k\cot(k\ell_j), \qquad B_j(k) := \frac{k}{\sin(k\ell_j)}.
  \]
  Ordering the vertices with the center $\mv$ coming last, we obtain
  the all-vertex DtN map
    \begin{equation}\label{eq:star_all_vertex_dtn}
    \Lambda(k;\Vset)=
    \begin{pmatrix}
      A_1 & 0   & 0   & 0   & B_1\\
      0   & A_2 & 0   & 0   & B_2\\
      0   & 0   & A_3 & 0   & B_3\\
      0   & 0   & 0   & A_4 & B_4\\
      B_1 & B_2 & B_3 & B_4 & A_1{+}A_2{+}A_3{+}A_4
    \end{pmatrix}.
  \end{equation}

  Removing the Dirichlet vertices (rows and columns 3 and 4) we get
  the partitioned matrix
  \begin{equation}\label{eq:star_reduced_partitioned}
    \Lambda(k;\partial\Vset\sqcup\Iset)=
    \left(
      \begin{array}{c|c}
        \Lambda_{\partial\Vset\ \partial \Vset} & \Lambda_{\partial \Vset\ \Iset}\\
        \hline
        \Lambda_{\Iset\ \partial \Vset} & \Lambda_{\Iset\ \Iset}
      \end{array}
    \right)
    =
    \left(
      \begin{array}{cc|c}
        A_1 & 0   & B_1\\
        0   & A_2 & B_2\\
        \hline
        B_1 & B_2 & A_1{+}A_2{+}A_3{+}A_4
      \end{array}
    \right).
  \end{equation}
  
  Using equation \eqref{eq:dtn_schur_complement} we finally arrive to
  \begin{align}
    \label{eq:star_dtn_final}
    \Lambda(k;\partial \Vset)
    &=
      \begin{pmatrix}
        A_1 & 0\\
        0 & A_2
      \end{pmatrix}
            -
            \begin{pmatrix}
              B_1\\B_2
            \end{pmatrix}
    \Big(A_1{+}A_2{+}A_3{+}A_4\Big)^{-1}
    \begin{pmatrix}
      B_1 & B_2
    \end{pmatrix}
            =
            \begin{pmatrix}
              A_1-\dfrac{B_1^2}{A_1{+}A_2{+}A_3{+}A_4}
              &
              -\dfrac{B_1B_2}{A_1{+}A_2{+}A_3{+}A_4}
              \\[10pt]
              -\dfrac{B_1B_2}{A_1{+}A_2{+}A_3{+}A_4}
              &
              A_2-\dfrac{B_2^2}{A_1{+}A_2{+}A_3{+}A_4}
            \end{pmatrix}.
  \end{align}
\end{example}

\begin{example}[Tadpole with two tails]
  \label{ex:loop_graph}
  Consider the graph in Figure~\ref{fig:gloop_DtN}.  The boundary is
  $\partial \Vset = \{v_1\}$; the conditions are Neumann--Kirchhoff at $v_2$
  and $v_3$, and Dirichlet at $v_4$.
  \begin{figure}
    \centering
    \begin{tikzpicture}[scale=1.5]
      % Vertices
      \coordinate (V2) at (-1,0);
      \coordinate (V3) at (1,0);
      \coordinate (V1) at (-2.5,0);
      \coordinate (V4) at (2.5,0);
      
      % Edges
      \draw[->,thick] (V2) -- (V1)
      node[at end,below=0.1cm] {$v_1$}
      node[midway, above] {$\me_1$};
      \draw[thick] (V3) -- (V4) node[midway, above] {$\me_4$};
      
      % Loop edges
      \draw[thick] (V2) to[bend left=45] node[midway, above] {$\me_2$} (V3);
      \draw[thick] (V2) to[bend right=45] node[midway, below] {$\me_3$} (V3);
      
      % Nodes
      \draw[fill] (V2) circle (1.5pt) node[below=0.1cm] {$v_1$};
      \draw[fill] (V3) circle (1.5pt) node[below=0.1cm] {$v_2$};
      \draw[fill=white] (V4) circle (1.5pt) node[below=0.1cm] {$v_4$};
    \end{tikzpicture}
    \caption{A graph with an internal loop, considered in \Cref{ex:loop_graph}.}
    \label{fig:gloop_DtN}
  \end{figure}
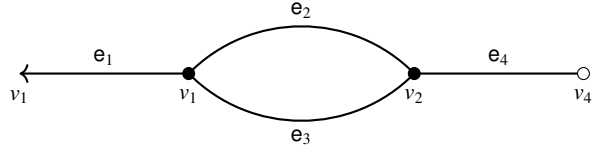

  Ordering the vertices as $(v_1,v_2,v_3,v_4)$, we obtain the all-vertex
  DtN map
  \begin{equation}\label{eq:loop_all_vertex_dtn}
    \Lambda(k;\Vset)=
    \begin{pmatrix}
      A_1 & B_1 & 0 & 0\\
      B_1 & A_1{+}A_2{+}A_3 & B_2{+}B_3 & 0\\
      0 & B_2{+}B_3 & A_2{+}A_3{+}A_4 & B_4\\
      0 & 0 & B_4 & A_4
    \end{pmatrix}.
  \end{equation}
  (The off-diagonal entry $B_2{+}B_3$ reflects the two parallel edges
  $\me_2,\me_3$ between $v_2$ and $v_3$.)

  Removing the Dirichlet vertex $v_4$ (i.e.\ deleting the last row and
  column), and partitioning with $\partial \Vset=\{v_1\}$ and
  $\Iset=\{v_2,v_3\}$, we obtain
  \begin{equation}\label{eq:loop_reduced_partitioned}
    \Lambda(k;\Vset\sqcup\Iset)=
    \left(
    \begin{array}{c|cc}
      \Lambda_{\partial \Vset\ \partial \Vset} & \Lambda_{\partial \Vset\ \Iset}\\
      \hline
      \Lambda_{\Iset\ \partial \Vset} & \Lambda_{\Iset\ \Iset}
    \end{array}
    \right)
    =
    \left(
    \begin{array}{c|cc}
      A_1 & B_1 & 0\\
      \hline
      B_1 & A_1{+}A_2{+}A_3 & B_2{+}B_3\\
      0   & B_2{+}B_3       & A_2{+}A_3{+}A_4
    \end{array}
    \right).
  \end{equation}
  (Note that $A_4$ remains in the $(v_3,v_3)$-entry since $v_4$ is
  Dirichlet, but the edge $\me_4$ still contributes to the Neumann
  data at $v_3$.)

  Using \eqref{eq:dtn_schur_complement} we arrive at the $1\times 1$
  DtN map at $v_1$:
  \begin{equation}\label{eq:loop_dtn_final}
    \Lambda(k;\Bset)
    = \begin{pmatrix}A_1\end{pmatrix}
    -
    \begin{pmatrix}B_1 & 0\end{pmatrix}
    \begin{pmatrix}
      A_1{+}A_2{+}A_3 & B_2{+}B_3\\
      B_2{+}B_3 & A_2{+}A_3{+}A_4
    \end{pmatrix}^{-1}
    \begin{pmatrix}B_1\\ 0\end{pmatrix}
    =
    A_1
    -\frac{B_1^2\,(A_2{+}A_3{+}A_4)}
    {(A_1{+}A_2{+}A_3)(A_2{+}A_3{+}A_4)-(B_2{+}B_3)^2}.
  \end{equation}

  Similarly to \eqref{eq:secular_dtn}, the zeros of the determinant $\det
  \Lambda(k;\partial \Vset)$ correspond to the eigenvalues of the original
  graph.  But they can miss some of the eigenvalues, for example those
  whose eigenfunction localizes on the loop (if $\ell_2=\ell_3$) and
  is identically zero on the edge $\me_1$.  This phenomenon is
  sometimes called ``lack of unique continuation'' in the mathematical
  literature and is closely connected with scars and topological
  resonances described in Section~\ref{sec:topological_resonances}.
\end{example}

\begin{example}[Square with two leads]
  \label{ex:square_DtN}
  Consider the square with 4 vertices numbered $v_1,v_3,v_2,v_4$
  clockwise, and with an additional diagonal edge $\me_5$ between
  $v_3$ and $v_4$, see \Cref{fig:gloop_DtN}.  The boundary set is
  $\partial \Vset=\{v_1,v_2\}$ and the conditions at $v_3,v_4$ are
  Neumann--Kirchhoff, see \Cref{fig:square_DtN}.
  
  \begin{figure}
    \centering
    \begin{tikzpicture}[scale=1.3]
      % Vertices (counter-clockwise: v1, v3, v2, v4)
      \coordinate (V1) at (-1,-1); % bottom-left
      \coordinate (V3) at (-1, 1); % top-left
      \coordinate (V2) at ( 1, 1); % top-right
      \coordinate (V4) at ( 1,-1); % bottom-right
      
      % Square edges
      \draw[thick] (V1) -- (V3) node[midway, left]  {$\me_1$};
      \draw[thick] (V3) -- (V2) node[midway, above] {$\me_2$};
      \draw[thick] (V2) -- (V4) node[midway, right] {$\me_3$};
      \draw[thick] (V4) -- (V1) node[midway, below] {$\me_4$};
      
      % Diagonal edge e5 between v3 and v4
      \draw[thick] (V3) -- (V4) node[midway, sloped, above] {$\me_5$};

      % "Boundary"
      \draw[->,thick] (V1) -- (-1.2,-1);
      \draw[->,thick] (V2) -- (1.2,1);
      
      % Boundary vertices (v1, v2): open circles
      \draw[fill] (V1) circle (1.5pt) node[below left=0.1cm] {$v_1$};
      \draw[fill] (V2) circle (1.5pt) node[above right=0.1cm] {$v_2$};
      
      % Interior vertices (v3, v4): filled dots
      \draw[fill] (V3) circle (1.5pt) node[above left=0.1cm] {$v_3$};
      \draw[fill] (V4) circle (1.5pt) node[below right=0.1cm] {$v_4$};
    \end{tikzpicture}
    \caption{A graph with two boundary vertices of degree 2.}
    \label{fig:square_DtN}
  \end{figure}
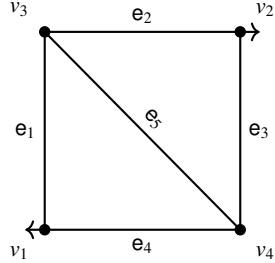

  The all-vertex DtN map, partitioned according to $\Vset=\partial\Vset\sqcup\Iset$ with
  $\Bset=\{v_1,v_2\}$ and $\Iset=\{v_3,v_4\}$, is
  \begin{equation}
    \label{eq:square_av_dtn}
    \Lambda(k;\Vset)=
    \left(
      \begin{array}{c|c}
        \Lambda_{\partial \Vset\ \partial \Vset} & \Lambda_{\partial \Vset\ \Iset}\\
        \hline
        \Lambda_{\Iset \ \partial \Vset} & \Lambda_{\Iset\ \Iset}
      \end{array}
    \right)
    =
    \left(
      \begin{array}{cc|cc}
        A_1{+}A_4 & 0 & B_1 & B_4\\
        0 & A_2{+}A_3 & B_2 & B_3\\
        \hline
        B_1 & B_2 & A_1{+}A_2{+}A_5 & B_5\\
        B_4 & B_3 & B_5 & A_3{+}A_4{+}A_5
      \end{array}
    \right).
  \end{equation}

  The Schur complement yields the DtN map 
  \begin{equation}\label{eq:square_dtn_explicit}
    \Lambda(k;\partial \Vset)
    =
    \begin{pmatrix}
      A_1{+}A_4 & 0\\
      0 & A_2{+}A_3
    \end{pmatrix}
    -
    \begin{pmatrix}
      B_1 & B_4\\
      B_2 & B_3
    \end{pmatrix}
    \begin{pmatrix}
      A_1{+}A_2{+}A_5 & B_5\\
      B_5 & A_3{+}A_4{+}A_5
    \end{pmatrix}^{-1}
    \begin{pmatrix}
      B_1 & B_2\\
      B_4 & B_3
    \end{pmatrix}.
  \end{equation}
\end{example}

\begin{remark}
  In the above examples, namely in equations
  \eqref{eq:star_all_vertex_dtn}, \eqref{eq:loop_all_vertex_dtn}, and
  \eqref{eq:square_av_dtn}, it is apparent that the all-vertex DtN map
  has the form of a generalized discrete Schr\"odinger operator
  corresponding to the graph, but with entries depending on the
  spectral parameter $k=\sqrt{\lambda}$.
\end{remark}

%%%%%%%%%%%%%%%%%%%%%%%%%%%%%%%%%%%%%%%%%%%%%%%%
\subsection{Spectral estimates from graph modifications}
\label{sec:spec_estimates}

In some applications, such as dynamics and stability of solutions to
nonlinear equations on graphs \cite{kairzhan2022standing}, it is
important to compare the eigenvalues of two graphs with same edge sets
but different connectivity.  Such comparison can often be achieved by
chaining together some well-studied graph modifications (``surgery
operations'') and the associated eigenvalue ineqalities.  We list some
of the results here, a comprehensive study is
\cite{BerKenKurMug_tams19}.

\subsubsection{Imposing Dirichlet conditions at vertices.}

The simplest operation is changing the conditions at one or more
vertices from $\delta$-type to Dirichlet.  Let $H$ be a self-adjoint
Hamiltonian on a compact metric graph and $\Vset^D$ be a
subset of its vertices; let $H^{D}$ be the
Hamiltonian obtained from $H$ by imposing Dirichlet conditions
$\psi(\mv)=0$ at all $\mv\in \Vset^D$ (leaving all other vertex conditions
unchanged).  Then the ordered eigenvalues of the two Hamiltonians
follow the inequality
\begin{equation}
  \label{eq:dirichlet_bracketing}
  \lambda_n(H) \leq \lambda_n \big(H^{D}\big)
  \leq \lambda_{n+d} (H),
  \qquad n=1,2,\dots,
\end{equation}
where $d = |\Vset^D|$ is the number of vertices where the Dirichlet
condition has been imposed.

The easiest way to prove \eqref{eq:dirichlet_bracketing} is to
consider the quadratic form (``energy form'') of the Hamiltonian $H$,
see \eqref{eq:energy_form},
whose domain imposes the continuity conditions on the function $\psi$
but, crucially, not on the derivative of $\psi$ (the conditions on the
derivative arise from the variational principle: the eigenfunctions
are the critical points of the energy form).  The domain of $H^D$ has
$d$ \emph{extra} conditions: the function is not only continuous but
is also equal to 0 at each of the $d$ vertices in $\Vset^D$.  Minimax
characterization of the eigenvalues implies that imposing extra
constraints pushes the eigenvalues higher, but no more ``places'' than
the extra number of constraints imposed.

\subsubsection{Splitting an Neumann-Kirchhoff vertex into $p$ vertices.}

Another basic surgery operation is \emph{splitting} a Neumann--Kirchhoff
vertex into several vertices.  Let $H$ be a self-adjoint Hamiltonian on
a compact metric graph and let $\mv$ be a vertex of degree
$\deg(\mv)=m$ at which $H$ satisfies the standard Neumann--Kirchhoff
(NK) conditions.  Fix an integer $p\in\{2,\dots,m\}$ and partition the
set of edges incident to $\mv$ into $p$ nonempty groups.  We form a new
graph by replacing $\mv$ with $p$ vertices $\mv^{(1)},\dots,\mv^{(p)}$,
and attaching to $\mv^{(j)}$ precisely the edges in the $j$th group.
At each of the new vertices $\mv^{(j)}$ we impose the Neumann-Kirchhoff
 conditions.
Denote by $H^{\mathrm{cut}}$ the resulting Hamiltonian.

Then the ordered eigenvalues satisfy the two-sided estimate
\begin{equation}
  \label{eq:cut_bracketing}
  \lambda_{n-(p-1)}(H)\ \le\ \lambda_n\big(H^{\mathrm{cut}}\big)
  \ \le\ \lambda_n(H),
  \qquad n=1,2,\dots,
\end{equation}
with the convention that $\lambda_k(H)=-\infty$ for $k\le 0$.
Equivalently, splitting an Neumann-Kirchhoff vertex can only \emph{lower} the
eigenvalues, but it cannot lower them by more than $(p-1)$ ``places''.

The proof is again most transparent on the level of energy forms, see
\eqref{eq:energy_form}.  The energy form is the same in both cases (it
is the sum of $\int | \psi'|^2$ over edges, plus possible potential
terms), but the form domain changes.  For the original operator $H$,
the domain requires that $\psi$ be continuous at $\mv$, i.e., that the
boundary values on all $m$ incident edges coincide at $\mv$.  After
splitting $\mv$ into $p$ vertices, continuity is only enforced
\emph{within each group} of edges, so we have removed exactly $(p-1)$
independent constraints on vertex values.  The minimax principle
implies that removing constraints enlarges the form domain and pushes
eigenvalues downwards.  The interlacing bound in
\eqref{eq:cut_bracketing} reflects that the number of removed
constraints is precisely $(p-1)$.

Eigenvalue estimates \eqref{eq:cut_bracketing} generalize
straightforwardly to the case when a $\delta$-type vertex is split
into several vertices (assuming the original coupling constant is the
sum of the new coupling constants).

\subsubsection{Increasing $\delta$-coupling constants at vertices.}

We next consider surgery operations which keep the underlying metric
graph fixed but change the $\delta$-type coupling strengths.  Let $H$
be a Hamiltonian with $\delta$-type conditions \eqref{eq:cond_current}
at all vertices (Dirichlet conditions being interpreted as $\alpha_\mv = +\infty$).
Let $\Vset^\uparrow\subset\Vset$
be a set of vertices with $d:=|\Vset^\uparrow|$, and define a second
Hamiltonian $\widetilde H$ by changing the couplings at these vertices
from $\alpha_{\mv}$ to $\widetilde\alpha_{\mv}$, while keeping all
other couplings the same:
\[
  \widetilde\alpha_{\mv}\ge \alpha_{\mv}\quad\text{for }\mv\in\Vset^\uparrow,
  \qquad
  \widetilde\alpha_{\mv}= \alpha_{\mv}\quad\text{for }\mv\notin\Vset^\uparrow.
\]
Then the eigenvalues satisfy the monotonicity (bracketing) inequalities
\begin{equation}
  \label{eq:delta_interlacing}
  \lambda_n(H)\ \le\ \lambda_n(\widetilde H)\ \le\ \lambda_{n+d}(H),
  \qquad n=1,2,\dots.
\end{equation}
Informally, strengthening the coupling at $d$ vertices pushes the
eigenvalues higher, but by no more than $d$ ``places''.

The proof again uses quadratic forms, see \eqref{eq:energy_form}.
Increasing $\alpha_{\mv}$ adds a nonnegative term
$(\widetilde\alpha_{\mv}-\alpha_{\mv})|\psi(\mv)|^2$ at each
$\mv\in\Vset^\uparrow$, hence
$\mathfrak{h}_{\widetilde\alpha}[\psi]\ge \mathfrak{h}_{\alpha}[\psi]$
for all admissible $\psi$, which yields the lower estimate in
\eqref{eq:delta_interlacing} by the minimax principle.  The estimate
from above reflects that the perturbation is supported on a
$d$-dimensional space of vertex values.

\section{Conclusion}

We have given an introduction to quantum graphs 
that should allow the reader to follow a large part of the
current literature using quantum graph models.
We have also given an overview over some recent applications in 
quantum chaos
and spectral theory that we find interesting. 
While we have given plenty of further in-depth reading in the corresponding chapters
we are aware that there are plenty of further interesting topics and results  
that we have not covered here. The textbooks and reviews we mentioned in the
introduction contain overviews and references for many of the topics we did not cover here.

\begin{ack}[Acknowledgments]\ We want to thank Uzy Smilansky who
  introduced us to the topic, and mentored us when we were both
  postdocs in his group at the same time.  Since then, we have had
  innumerable discussions with him about quantum graphs and their
  applications to quantum chaos and spectral theory.  We are grateful
  to Ram Band, Pavel Exner, James Kennedy, Peter Kuchment, Delio
  Mugnolo and Brian Winn for their suggestions of important
  references.  Jon Harrison kindly pointed out several typos and
  inconsistencies in a preliminary version of the manuscript.  A
  discussion with Michael Levitin has inspired us to develop a
  practical guide to the computation of the Dirichlet-to-Neumann maps
  in Section~\ref{sec:dtn}.
\end{ack}

%\seealso{article title article title}

\providecommand{\href}[2]{#2}\begingroup\raggedright\endgroup

%\bibliographystyle{JHEP}%
%\bibliographystyle{siam}
%\bibliography{MRW-CQUA-reference,bk_bibl}
%\bibliography{bk_bibl,qc-bibliography}
\end{document}